\documentclass[preprintnumbers,showpacs,showkeys,preprint,secnumarabic,amssymb,amsmath,amsfonts,aps]{revtex4-1}

\usepackage{amsmath}
\usepackage{amsfonts}

\usepackage{yfonts}
\usepackage{mathrsfs}

\usepackage{amssymb}
\usepackage[english]{babel}
\usepackage[pdftex]{graphicx}
\usepackage{tikz}
\usepackage{rotating}
\usepackage[pdftex,hypertexnames=true,linktocpage=true,breaklinks=true]{hyperref} 
\usepackage[hyphenbreaks]{breakurl}
\hypersetup{colorlinks=true,linkcolor=blue,anchorcolor=blue,citecolor=magenta,filecolor=blue,urlcolor=blue,bookmarksnumbered=false,pdfview=FitB}

\newcommand{\bea}{\begin{eqnarray}}
\newcommand{\eea}{\end{eqnarray}}
\newcommand{\bml}{\begin{mathletters}}
\newcommand{\eml}{\end{mathletters}}

\newcommand{\beq}{\begin{equation}}
\newcommand{\eeq}{\end{equation}}
\newcommand{\affF}{DSFTA, University of Siena, Via Roma 56, 53100 Siena, Italy}

\newcommand{\affA}{Aix-Marseille University, Marseille, France}
\newcommand{\affB}{CNRS Centre de Physique Th\'eorique UMR7332,
13288 Marseille, France}

\newcommand{\affD}{Dipartimento di Fisica Universit\`a di Firenze, and
I.N.F.N., Sezione di Firenze, via G. Sansone 1, I-50019 Sesto Fiorentino, Italy }
\newcommand{\affN}{Department of Physics and Sciences of Materials, University of Luxembourg, Luxembourg}
\newcommand{\affM}{Quantum Biology Lab, Howard University, 2400 6th St NW, Washington, DC 20059, USA }
\everymath{\displaystyle}
\begin{document}

\title{Hamiltonian Dynamics and Fundamental Phenomena in Biophysics: A review}

\author{Matteo Gori}
\email{gori6matteo@gmail.com}
\affiliation{\affN}

\author{Roberto Franzosi}
\email{roberto.franzosi@ino.it}
\affiliation{\affF}

\author{Giulio Pettini}
\email{pettini@fi.unifi.it}
\affiliation{\affD}

\author{Marco Pettini}
\email{pettini@cpt.univ-mrs.fr}
\affiliation{\affA}\affiliation{\affB}\affiliation{\affM}

\date{\today}

\begin{abstract}
We review a theoretical and experimental programme aimed at understanding 
two intimately related fundamental phenomena in biophysics:
(i)~the classical analogue of Fr\"ohlich phonon condensation in macromolecules
driven out of thermal equilibrium, and
(ii)~the consequent activation of long-range resonant electrodynamic
intermolecular forces.
Both phenomena are underpinned by explicit Hamiltonian models.
The first is derived by applying the time-dependent variational principle
(TDVP) to the quantum Wu--Austin model, producing a fully classical
Hamiltonian in action-angle variables whose nonlinear rate equations
exhibit a nonequilibrium phase transition, the channelling of
supplied energy into the lowest-frequency collective mode.
The second is grounded in a classical electrodynamic Hamiltonian
for two coupled oscillating dipoles whose normal-mode structure
predicts long-range ($\sim 1/r^3$) resonant interactions, absent
at thermal equilibrium but activated by out-of-equilibrium collective
oscillations.
We also discuss a complementary Hamiltonian approach that connects
Fr\"ohlich's rate equations directly to Hamilton's equations of motion,
clarifying the role of bath-mediated nonlinear coupling and the
conditions for strong condensation at room temperature.
In addition, the TDVP is applied to a Davydov--Holstein--Fr\"ohlich
Hamiltonian describing electron--phonon motion along the backbone of a
specific DNA sequence and its cognate restriction enzyme EcoRI:
the time-domain Fourier cross-spectrum of the resulting electron currents
exhibits a sharp co-resonance peak for the canonical recognition sequence
that disappears upon randomisation, providing a sequence-specific
electrodynamic signature of DNA--protein recognition.
Experimental evidence from THz near-field spectroscopy, fluorescence
correlation spectroscopy, and direct observation of protein clustering
is reviewed in relation to these theoretical predictions.
The results establish a coherent physical picture suggesting that metabolic
energy supply can play a role in driving macromolecules into coherently oscillating states
that activate selective, distance-reaching electrodynamic forces
capable of contributing to the organisation of biochemical reactions
in living matter.
\end{abstract}

\maketitle

%============================================================
\section{Introduction}
\label{sec:intro}
%============================================================

A fundamental challenge in biophysics is to understand how biochemical
reactions in living cells proceed with the speed and selectivity observed
experimentally, given the crowded, thermally noisy environment of the
cytoplasm.
Standard models invoke short-range forces, electrostatic, van der Waals,
hydrophobic, and diffusion-driven random encounters.
These mechanisms are undoubtedly operative, but they do not obviously account
for the remarkably high efficiency and rapidity with which cognate molecular
partners find one another among thousands of competing species in a
sub-micron space~\cite{Banani2017,Berry2018,Sweetlove2018}.

A minority but persistent view in theoretical biophysics, originating with
H.~Fr\"ohlich in the late
1960s~\cite{Frohlich1968,Frohlich1970,Frohlich1972,Frohlich1977},
proposes that biological macromolecules driven out of thermodynamic
equilibrium by metabolic energy could undergo a collective ordering
transition, a \emph{phonon condensation}, in which the input energy
concentrates into the lowest-frequency vibrational mode of a macromolecule,
producing a coherent giant oscillating dipole moment.
This coherent oscillation was predicted to activate resonant, selective,
long-range electrodynamic forces between macromolecules vibrating at the same
frequency~\cite{Frohlich1972}, forces that could accelerate and direct
molecular recognition processes.

Three key obstacles have historically limited the impact of this programme.
First, the original quantum Wu--Austin microscopic Hamiltonian~\cite{Wu1977,Wu1978}
suffers from an instability that requires stabilisation.
Second, the connection between the rate-equation level and direct Hamiltonian
dynamics had not been established in a classical framework accessible to
modern molecular dynamics.
Third, and most critically, no experimental evidence for out-of-equilibrium
collective oscillations of macromolecules existed until recently.

All these obstacles have now been overcome.
The present work reviews the following pivotal topics: 
\textit{i)} A classical Hamiltonian in action-angle variables is derived from the
      quantum Wu--Austin model via the TDVP applied to product coherent states
      (Sec.~\ref{sec:WA}--\ref{sec:condensation}).
      Nonlinear rate equations follow from the Koopman--von Neumann (KvN)
      reformulation of the Liouville equation.
      Their stationary solutions exhibit classical phonon condensation~\cite{Nardecchia2018}.
\textit{ii)} classical electrodynamic Hamiltonian for two coupled oscillating
      dipoles is analysed via normal-mode theory (Sec.~\ref{sec:EDforces}).
      The result is a long-range ($\sim 1/r^3$) interaction potential active
      only out of thermal equilibrium~\cite{Preto2015}.
\textit{iii)} A position-space classical Hamiltonian dynamics, closer to standard molecular
      dynamics, is simulated directly (Sec.~\ref{sec:hamMD}) recovering the outcomes of 
      the Fr\"ohlich rate equations, and showing the formation of strong
      condensates at room temperature are found under resonant bath-mediated
      coupling~\cite{Preto2024}.
\textit{iv)} The TDVP is applied to a Davydov--Holstein--Fr\"ohlich Hamiltonian for
      electron--phonon motion along a specific DNA sequence and its cognate restriction 
      enzyme EcoRI (Sec.~\ref{sec:DNA}), and the associate dynamics yields 
      a sharp cross-spectral co-resonance between the electron currents of
      the two molecules appears for the canonical recognition sequence and
      disappears upon randomisation~\cite{Faraji2025}.

Experimental evidence is reviewed in Sec.~\ref{sec:experiment}.
Sec.~\ref{sec:outlook} gives an outlook of future developments.

%============================================================
\section{The Wu--Austin Quantum Model and the Classical Limit}
\label{sec:WA}
%============================================================
\subsection{Physical picture}

Fr\"ohlich~\cite{Frohlich1968,Frohlich1970} modelled a macromolecule as an
open system: an ensemble of coupled vibrational modes in contact simultaneously
with a thermal bath at temperature $T_B$ and with an energy source at
effective temperature $T_S \gg T_B$.
He argued that when the energy-input rate exceeds a threshold, the system
undergoes a nonequilibrium phase transition: instead of distributing the
supplied energy among all modes (equipartition), it channels nearly all of
it into the lowest-frequency mode, producing a macroscopic coherent
oscillation.

\subsubsection{Fr\"ohlich rate equations and condensation}

At the phenomenological level, Fr\"ohlich described the dynamics in terms of
the occupation numbers $n_k(t)=\langle\hat{a}^\dagger_{\omega_k}\hat{a}_{\omega_k}\rangle$
of the vibrational modes~\cite{Frohlich1977}.
Consider a set of $z$ polar modes with frequencies
$\omega_1\leq\omega_k\leq\omega_z$, $k=1,\dots,z$, coupled to a heat bath
at inverse temperature $\beta_B=1/k_BT_B$ and driven by an external source
that injects quanta into mode $k$ at rate $s_k$.
Two classes of bath-mediated processes are retained.

\emph{First-order} processes exchange a single quantum between mode $k$ and
the bath, tending to restore thermal equilibrium.
\emph{Second-order} processes transfer excitations between two different
system modes $k$ and $j$ with the bath absorbing or supplying the energy
difference; crucially, they conserve the total number of system quanta,
$\sum_k\dot{n}_k^{(2)}=0$.
Denoting by $\varphi$ the linear relaxation coefficient and by $\chi$ the
nonlinear intermode coupling, the complete rate equations are (for $k=1,\dots,z$)
\begin{equation}
  \dot{n}_k = s_k
    - \varphi\!\left[n_k\,e^{\beta_B\hbar\omega_k}-(n_k+1)\right]
    - \chi\sum_{j\neq k}\!\left[
        n_k(1+n_j)\,e^{\beta_B\hbar\omega_k}
        -n_j(1+n_k)\,e^{\beta_B\hbar\omega_j}
      \right].
  \label{eq:FrohlichRate}
\end{equation}
The first term injects energy from the source; the second drives each mode
toward thermal equilibrium with the bath; the third redistributes quanta
across the spectrum without altering their total number.

\paragraph{Effective chemical potential.}
This nonlinear redistribution is the key to the Fr\"ohlich effect.
Because downward transitions in frequency are statistically favoured by the
Bose factors $(1+n_j)$, pumped energy is not simply equipartitioned.
Once the total input rate $S=\sum_k s_k$ exceeds a threshold set by the
competition between pumping, dissipation, and nonlinear scattering, the
excess population can no longer be absorbed by the higher-frequency modes
and accumulates at the bottom of the spectrum.
In the stationary regime one introduces an effective nonequilibrium chemical
potential $\mu$, defined self-consistently by
\begin{equation}
  e^{-\beta_B\mu}
  = \frac{\varphi+\chi\displaystyle\sum_j(1+n_j)}
         {\varphi+\chi\displaystyle\sum_j n_j\,e^{\beta_B\hbar\omega_j}},
  \label{eq:mu_def}
\end{equation}
which satisfies $0\leq\mu<\hbar\omega_1$ whenever $S>0$.
The formal stationary solution then takes the Bose-like form
\begin{equation}
  n_k = \left(1+\frac{\varphi}{\chi}\frac{s_k}{S}
              \bigl(1-e^{-\beta_B\mu}\bigr)\right)
        \frac{1}{e^{\beta_B(\hbar\omega_k-\mu)}-1},
  \label{eq:nk_sol}
\end{equation}
so that as $S$ increases, $\mu$ approaches $\hbar\omega_1$ from below and
$n_1$ grows without bound while all other occupations remain comparatively
limited.

\paragraph{Fr\"ohlich condensation.}
This is the Fr\"ohlich condensation phenomenon: a nonequilibrium ordered
state in which a single low-frequency polar mode becomes macroscopically
populated and behaves as a collective coherent oscillation of the
macromolecule.
The analogy with Bose--Einstein condensation is instructive: in both cases
the discrete sum over modes cannot be replaced by an integral once $\mu$
saturates at the band edge, forcing a macroscopic occupation of the ground
mode.
The mechanism is, however, intrinsically a nonequilibrium one.
In an equilibrium Bose gas the condensed state is reached by lowering the
temperature; here it is reached by maintaining a sufficiently strong energy
throughput in an open, dissipative system at fixed $T_B$.
When the nonlinear coupling vanishes ($\chi=0$), equation~\eqref{eq:mu_def}
gives $\mu=0$ and \eqref{eq:nk_sol} reduces to a modified Planck
distribution with no condensation; it is therefore the nonlinear
bath-mediated intermode scattering that is the essential ingredient.

The predicted coherent oscillation lies in the sub-THz range,
$\nu\sim 10^{11}$~Hz, set by acoustic-like breathing vibrations of proteins
or membrane sections of linear size $\ell\sim 10^{-6}$~cm with effective
sound velocity $v\sim 10^5$~cm/s~\cite{Frohlich1977}.
Once the lowest polar mode is macroscopically occupied it acts as an
oscillating giant dipole; cell water with small ions in solution, with typical Debye length of 10\AA, screens static charges
efficiently, but does not screen oscillating fields at these frequencies.
Two such coherently oscillating dipoles at separation $R$ interact via a
frequency-selective, attractive potential that in the near-resonance
limit falls off as~\cite{Frohlich1977}
\begin{equation}
  I \simeq -U\,\frac{e^2(z_e z_s)^{1/2}}{M\omega_e^2\,R^3}
  \;\propto\; -\frac{1}{R^3},
  \label{eq:Frohlich_LR}
\end{equation}
where $U=\hbar n_-\omega_e$ is the excitation energy, $z_e,z_s$ the numbers
of coherently oscillating charges, $M$ their mass, and $n_-$ the occupation
of the lower normal mode of the coupled pair.
This long-range force is proposed as the mechanism for frequency-selective
recognition between enzymes and their substrates and for the attraction of
homologous chromosomes during meiosis~\cite{Frohlich1977}.

\subsubsection{Microscopic derivation: Wu--Austin Hamiltonian}

A microscopic quantum Hamiltonian from which Fr\"ohlich's rate equations
can be derived was given by Wu and Austin~\cite{Wu1977,Wu1978}.
The system consists of three sets of quantum harmonic oscillators:
normal modes of the macromolecule (frequencies $\omega_i \in I_{\rm sys}$,
operators $\hat{a}_{\omega_i}$, $\hat{a}^\dagger_{\omega_i}$);
a thermal bath at $T_B$ (frequencies $\Omega_j \in I_{\rm bth}$, operators
$\hat{b}_{\Omega_j}$, $\hat{b}^\dagger_{\Omega_j}$);
and an energy source modelled as a second bath at $T_S \gg T_B$
(frequencies $\Omega'_k \in I_{\rm src}$, operators $\hat{c}_{\Omega'_k}$,
$\hat{c}^\dagger_{\Omega'_k}$).
The total Hamiltonian is $\hat{H}_{\rm Tot} = \hat{H}_0 + \hat{H}_{\rm int}$,
where
\begin{equation}
  \hat{H}_0 = \sum_{\omega_i}\hbar\omega_i\,\hat{a}^\dagger_{\omega_i}\hat{a}_{\omega_i}
            + \sum_{\Omega_j}\hbar\Omega_j\,\hat{b}^\dagger_{\Omega_j}\hat{b}_{\Omega_j}
            + \sum_{\Omega'_k}\hbar\Omega'_k\,\hat{c}^\dagger_{\Omega'_k}\hat{c}_{\Omega'_k},
  \label{eq:H0quant}
\end{equation}
the interaction contains the Wu--Austin trilinear couplings
\begin{equation}
  \hat{H}_{\rm int}^{\rm WA}
  = \sum_{\omega_i,\Omega_j}
       \eta_{\omega_i\Omega_j}\,\hat{a}^\dagger_{\omega_i}\hat{b}_{\Omega_j}
   + \sum_{\omega_i,\Omega'_k}
       \xi_{\omega_i\Omega'_k}\,\hat{a}^\dagger_{\omega_i}\hat{c}_{\Omega'_k}
   + \sum_{\omega_i,\omega_j,\Omega_k}
       \chi_{\omega_i\omega_j\Omega_k}\,
       \hat{a}^\dagger_{\omega_i}\hat{a}_{\omega_j}\hat{b}^\dagger_{\Omega_k}
   + \mathrm{H.c.},
  \label{eq:HWA}
\end{equation}
supplemented by a stabilising quartic term~\cite{Bolterauer1999}
\begin{equation}
  \hat{H}_{\rm int}^{Q}
  = \sum_{\omega_i,\omega_j,\omega_k,\omega_l}
    \Bigl[
      \kappa^{(1)}_{\omega_i\omega_j\omega_k\omega_l}\,
        \hat{a}^\dagger_{\omega_i}\hat{a}^\dagger_{\omega_j}
        \hat{a}_{\omega_k}\hat{a}_{\omega_l}
    + \kappa^{(2)}_{\omega_i\omega_j\omega_k\omega_l}\,
        \hat{a}^\dagger_{\omega_i}\hat{a}^\dagger_{\omega_j}
        \hat{a}^\dagger_{\omega_k}\hat{a}_{\omega_l}
    + \mathrm{H.c.}
    \Bigr].
  \label{eq:HQ}
\end{equation}
The quartic term removes the pathology of the pure Wu--Austin model and
corresponds physically to anharmonic interactions among the macromolecular
modes.
Fr\"ohlich's phenomenological rate equations~\eqref{eq:FrohlichRate} are
recovered from this Hamiltonian by applying the quantum Liouville--von~Neumann
equation $i\hbar\,\partial_t\hat{\rho}=[\hat{H}_{\rm Tot},\hat{\rho}]$ to
the full density matrix $\hat{\rho}$, tracing over the bath and source
degrees of freedom in the Born--Markov approximation, and identifying
$n_k=\mathrm{Tr}\!\left(\hat{a}^\dagger_{\omega_k}\hat{a}_{\omega_k}\,
\hat{\rho}_{\rm sys}\right)$.
The trilinear terms in $\hat{H}_{\rm int}^{\rm WA}$ generate the linear
relaxation ($\varphi$) and the pumping ($s_k$), while the trilinear
system--bath coupling $\chi_{\omega_i\omega_j\Omega_k}$ produces the
nonlinear intermode redistribution term ($\chi$) in~\eqref{eq:FrohlichRate};
the quartic term $\hat{H}_{\rm int}^Q$ ensures the thermodynamic stability
of the condensed state.
\subsubsection{Why a classical description suffices}
At room temperature ($T \approx 300$\,K) and at the sub-THz frequencies
relevant to collective protein vibrations ($\omega/2\pi \sim 0.1$--$1$\,THz),
$k_BT/\hbar\omega \gg 1$, so the field is in the classical regime and the
Bose--Einstein mean occupation number $\langle n \rangle_{\rm BE}$ is well
approximated by the Boltzmann estimate $k_BT/\hbar\omega$, ranging from
$\approx 6$ at $1$\,THz to $\approx 62$ at $0.1$\,THz.
When driven out of equilibrium the effective temperature of each mode
increases further, making the classical approximation even better.
A classical framework is therefore not only legitimate but natural, and it
connects directly to the language of molecular dynamics simulation.
%============================================================
\section{Dequantization via the Time-Dependent Variational Principle}
\label{sec:TDVP}
%============================================================
%\subsection{The variational principle and its action}
The time-dependent variational principle
(TDVP)~\cite{Kramer1980,Kramer2008,Jauslin2010} extracts a classical
Hamiltonian from a quantum one by restricting dynamics to a
finite-dimensional manifold of trial states $|\psi(x_1,\ldots,x_N)\rangle$
parametrised by real variables $\{x_i(t)\}$.
The equations of motion follow from
\begin{equation}
  \delta S = 0, \qquad
  S = \int_0^t L(\psi,\bar{\psi})\,dt',
  \label{eq:TDVP}
\end{equation}
with the Lagrangian
\begin{equation}
  L = i\hbar\,
      \frac{\langle\psi|\dot\psi\rangle - \langle\dot\psi|\psi\rangle}
           {2\langle\psi|\psi\rangle}
    - \frac{\langle\psi|\hat{H}_{\rm Tot}|\psi\rangle}{\langle\psi|\psi\rangle}.
  \label{eq:Lagrangian}
\end{equation}
The resulting classical Hamiltonian is simply the expectation value
$H_{\rm Tot} = \langle\psi|\hat{H}_{\rm Tot}|\psi\rangle$.
This procedure has been advocated as a ``dequantization'' method, a kind
of inverse of geometric quantization~\cite{Jauslin2010}.
%\subsubsection{Coherent-state parametrisation}
Since $\hat{H}_{\rm Tot}$ is built from bosonic creation and annihilation
operators, the natural trial states are products of coherent states over all
modes in $I_S = I_{\rm sys} \cup I_{\rm bth} \cup I_{\rm src}$:
\begin{equation}
  |\Psi(t)\rangle =
    \bigotimes_{\omega_i \in I_{\rm sys}} |z_{\omega_i}(t)\rangle
    \otimes
    \bigotimes_{\Omega_j \in I_{\rm bth}} |z_{\Omega_j}(t)\rangle
    \otimes
    \bigotimes_{\Omega'_k \in I_{\rm src}} |z_{\Omega'_k}(t)\rangle,
  \label{eq:ansatz}
\end{equation}
where $|z\rangle = e^{-|z|^2/2}\sum_{k=0}^\infty z^k/\sqrt{k!}\,|k\rangle$
satisfies $\langle\hat{n}\rangle = |z|^2$.
%\subsection{Action-angle variables and canonical structure}
Writing $z_i = n_i^{1/2} e^{-i\theta_i}$ introduces real pairs
$(n_i, \theta_i)$ for every mode.
A direct computation of the symplectic tensor
$W_{ij} = \partial_i w_j - \partial_j w_i$, with
$w_i = i\hbar\langle\psi|\partial_{x_i}\psi\rangle$, yields
\begin{equation}
  W_{n_i\theta_k} = -W_{\theta_k n_i} = \hbar\,\delta_{ik},
  \qquad
  W_{\theta_i\theta_k} = W_{n_in_k} = 0.
  \label{eq:symplectic}
\end{equation}
Hence $J_{\omega_i} \equiv \hbar n_{\omega_i}$ and $\theta_{\omega_i}$ are
\emph{canonically conjugated} variables, $\{J_{\omega_i},\,\theta_{\omega_k}\} = \delta_{ik}$.
The action $J_\omega = \hbar\langle\hat{n}_\omega\rangle$ equals $\hbar$ times
the mean occupation number.
%---------------------------------------------------------------------------------------
\subsection{The classical Hamiltonian in action-angle variables}
Evaluating
$H_{\rm Tot} = \langle\Psi|\hat{H}_{\rm Tot}|\Psi\rangle$
in $(J_\omega,\theta_\omega)$ variables one finds~\cite{Nardecchia2018}:
\begin{align}
  H_{\rm Tot}
  &= \sum_{\omega_i} \omega_i J_{\omega_i}
   + \sum_{\Omega_j} \Omega_j J_{\Omega_j}
   + \sum_{\Omega'_k} \Omega'_k J_{\Omega'_k} 
   + \sum_{\omega_i,\Omega_j}
       \eta_{\omega_i\Omega_j}
       J_{\omega_i}^{1/2} J_{\Omega_j}^{1/2}
       \cos(\theta_{\omega_i} - \theta_{\Omega_j})
  \nonumber\\
  &\quad
   + \sum_{\omega_i,\Omega'_k}
       \xi_{\omega_i\Omega'_k}
       J_{\omega_i}^{1/2} J_{\Omega'_k}^{1/2}
       \cos(\theta_{\omega_i} - \theta_{\Omega'_k}) 
   + \sum_{\omega_i,\omega_j,\Omega_k}
       \chi_{\omega_i\omega_j\Omega_k}
       J_{\omega_i}^{1/2} J_{\omega_j}^{1/2} J_{\Omega_k}^{1/2}
       \cos(\theta_{\omega_i} - \theta_{\omega_j} + \theta_{\Omega_k})
  \nonumber\\
  &\quad
   + \sum_{\omega_i,\omega_j,\omega_k,\omega_l}
       J_{\omega_i}^{1/2} J_{\omega_j}^{1/2} J_{\omega_k}^{1/2} J_{\omega_l}^{1/2}
  \times\bigl[ \kappa^{(1)}_{\cdots}
           \cos(\theta_{\omega_i}+\theta_{\omega_j}-\theta_{\omega_k}-\theta_{\omega_l})
  + \kappa^{(2)}_{\cdots} \cos(\theta_{\omega_i}+\theta_{\omega_j}+\theta_{\omega_k}-\theta_{\omega_l}) \bigr].
  \label{eq:Hclass}
\end{align}
This is a fully classical, canonical Hamiltonian retaining the complete
structure of the quantum model.
Hamilton's equations
$\dot{J}_\omega = -\partial H/\partial\theta_\omega$, \, 
$\dot{\theta}_\omega = \partial H/\partial J_\omega$
govern all degrees of freedom.
%============================================================
\subsection{Classical Rate Equations via the Koopman--von Neumann Formalism}
\label{sec:rate}

Rather than integrating Hamilton's equations with an explicit bath, one works
with the phase-space probability density
$\rho(\{(J_\omega,\theta_\omega)\}; t)$ satisfying the Liouville equation
\begin{equation}
  \frac{\partial\rho}{\partial t}
  = \{H_{\rm Tot},\,\rho\}
  \equiv -i\hat{\mathcal{L}}_H\rho.
  \label{eq:Liouville}
\end{equation}
The Koopman--von Neumann (KvN) formalism~\cite{Koopman1931,vonNeumann1932}
rewrites this as a Schr\"odinger-like equation
$i\,\partial_t\psi = \hat{\mathcal{L}}_H\psi$, $\rho = |\psi|^2$,
on the Hilbert space $L^2(\Lambda)$, allowing interaction-picture perturbation
theory to be applied directly to the classical problem. In this framework the
action variables $J_{\omega_i} = \hbar n_{\omega_i}$ play the role of
classical counterparts of the quantum occupation numbers, and the conjugate
angles $\theta_{\omega_i}$ are the corresponding phases. Decomposing
$\hat{\mathcal{L}}_H = \hat{\mathcal{L}}_{H_0} + \hat{\mathcal{L}}_{H_{\rm int}}$
and treating $\hat{\mathcal{L}}_{H_{\rm int}}$ as a time-dependent perturbation
adiabatically switched on from $t_0=0$, one works in the interaction picture
and expands to second order. Averaging over thermal-bath initial conditions
compatible with temperatures $T_B$ (bath) and $T_S \gg T_B$ (external source)
then yields closed rate equations for $\langle J_{\omega_i}\rangle$, in full
analogy with the quantum derivation of Fröhlich's equations via time-dependent
perturbation theory~\cite{Wu1977,Wu1978}.

\subsection{The Classical Fr\"ohlich-like Rate Equations}
\label{sec:frohlich_rate}

The result of the second-order KvN expansion~\cite{Nardecchia2018} is
\begin{equation}
  \frac{d\langle J_{\omega_i}\rangle}{dt}
  = s_{\omega_i}
   + b_{\omega_i}\!\left(
       \frac{k_B T_B}{\omega_i} - \langle J_{\omega_i}\rangle
     \right)
   + \sum_{\omega_j \neq \omega_i}
       c_{\omega_i\omega_j}
       \Bigl[
         \langle J_{\omega_j}\rangle - \langle J_{\omega_i}\rangle
         + \frac{\omega_j - \omega_i}{k_B T_B}
           \langle J_{\omega_i}\rangle\langle J_{\omega_j}\rangle
       \Bigr]
   + (\text{quartic}),
  \label{eq:rate}
\end{equation}
where $s_{\omega_i} > 0$ is the energy-input rate from the external source,
$b_{\omega_i}$ is the linear system-bath coupling constant (with characteristic
thermalization timescale $\tau_{\rm therm} \approx b_{\omega_i}^{-1}$), and
$c_{\omega_i\omega_j}$ encodes the nonlinear mode-mode coupling mediated by the
bath. The quartic correction terms, absent in the original Wu--Austin model but
necessary to bound the energy from below~\cite{Bolterauer1999}, involve the
coupling constants $\kappa^{(1,2,3)}_{\omega_i\omega_j\omega_k\omega_l}$
of the anharmonic interaction among normal modes~\cite{Nardecchia2018}.

This equation has the following key properties. At equilibrium
($s_{\omega_i}=0$) the unique stationary solution is
$\langle J_{\omega_i}\rangle = k_B T_B/\omega_i$ (energy equipartition),
since $y_{\omega_i} = \omega_i\langle J_{\omega_i}\rangle/(k_B T_B) = 1$
for all modes. Equation~\eqref{eq:rate} is functionally identical to
Fröhlich's original quantum rate equations, with $\langle J_{\omega_i}\rangle$
replacing the occupation number $\langle\hat{n}_{\omega_i}\rangle$; this
functional identity is the classical counterpart of the Bose-like condensation
mechanism. As $s_{\omega_i}$ is increased above a threshold value $s_c$, the
stationary solution bifurcates away from equipartition: an increasingly large
fraction of the total energy is channelled into the lowest-frequency mode
$\omega_0$, giving rise to classical phonon condensation. Numerical integration
of the dimensionless form of Eq.~\eqref{eq:rate} confirms that this threshold
sharpens with increasing number of modes $N$, approaching a sharp bifurcation
in the thermodynamic limit~\cite{Nardecchia2018}.

Introducing $\tau = t\omega_0$, $y_{\omega_i} =
\omega_i\langle J_{\omega_i}\rangle/(k_B T_B)$, $\alpha_{\omega_i} =
\omega_i/\omega_0$ where $\omega_0 = \min_\omega\omega$, the rate equations
take the dimensionless form
\begin{equation}
  \dot{y}_{\omega_i}
  = S_{\omega_i} + B_{\omega_i}(1 - y_{\omega_i})
   + \sum_{\omega_j \neq \omega_i} C_{\omega_i\omega_j}
     \Bigl[
       \frac{\alpha_{\omega_i}}{\alpha_{\omega_j}} y_{\omega_j} - y_{\omega_i}
       + \frac{\alpha_{\omega_j} - \alpha_{\omega_i}}{\alpha_{\omega_j}}
         y_{\omega_i} y_{\omega_j}
     \Bigr]
   + (\text{quartic}),
  \label{eq:ratend}
\end{equation}
with dimensionless control parameter $S = S_{\omega_i}$ (assumed
mode-independent). The equilibrium solution is uniformly $y_{\omega_i} = 1$;
condensation corresponds to a stationary solution with $y_{\omega_0} \gg 1$
and $y_{\omega_i} \ll y_{\omega_0}$ for $\omega_i > \omega_0$.
%============================================================
\section{Classical Phonon Condensation}
\label{sec:condensation}
%============================================================

\subsection{Stationary states and order parameters}

The stationary solutions ($\dot{y}_{\omega_i} = 0$) at small $S$ are the
equipartition fixed point $y_{\omega_i} = 1$, invariant under permutations of
mode labels.
Above a threshold $S_c$ the nonlinear coupling breaks this permutation
symmetry: energy flows preferentially toward $\omega_0$.

Two order parameters quantify this.
The condensation index~\cite{Reimers2009}
\begin{equation}
  E_y = \frac{y_{\omega_0}/\tilde{y}_{\omega_0}}
             {\sum_{i=0}^N y_{\omega_i}},
  \quad \tilde{y}_{\omega_0} = 1 + S/B,
  \label{eq:Ey}
\end{equation}
takes $E_y = 0$ at equipartition and $E_y = 1$ at full condensation.
The ratio $p_1/p_0$ (where $p_i = y_{\omega_i}/\sum_j y_{\omega_j}$)
goes from $1$ to $0$ at the same transition.

\subsection{Numerical evidence}

The rate equations are numerically integrated  for $N+1$
equally spaced modes $\omega_n = \omega_0(1 + n/N)$, starting from
$y_{\omega_i}(0) = 1$.
Parameters: $B = 1$, $C = 0.1$, quartic coupling
$\Upsilon^{(1)} = \Upsilon^{(2)} = 10^{-4}$~\cite{Nardecchia2018}.

\begin{figure}[h!]
 \centering
\includegraphics[scale=0.45,keepaspectratio=true,angle=0]{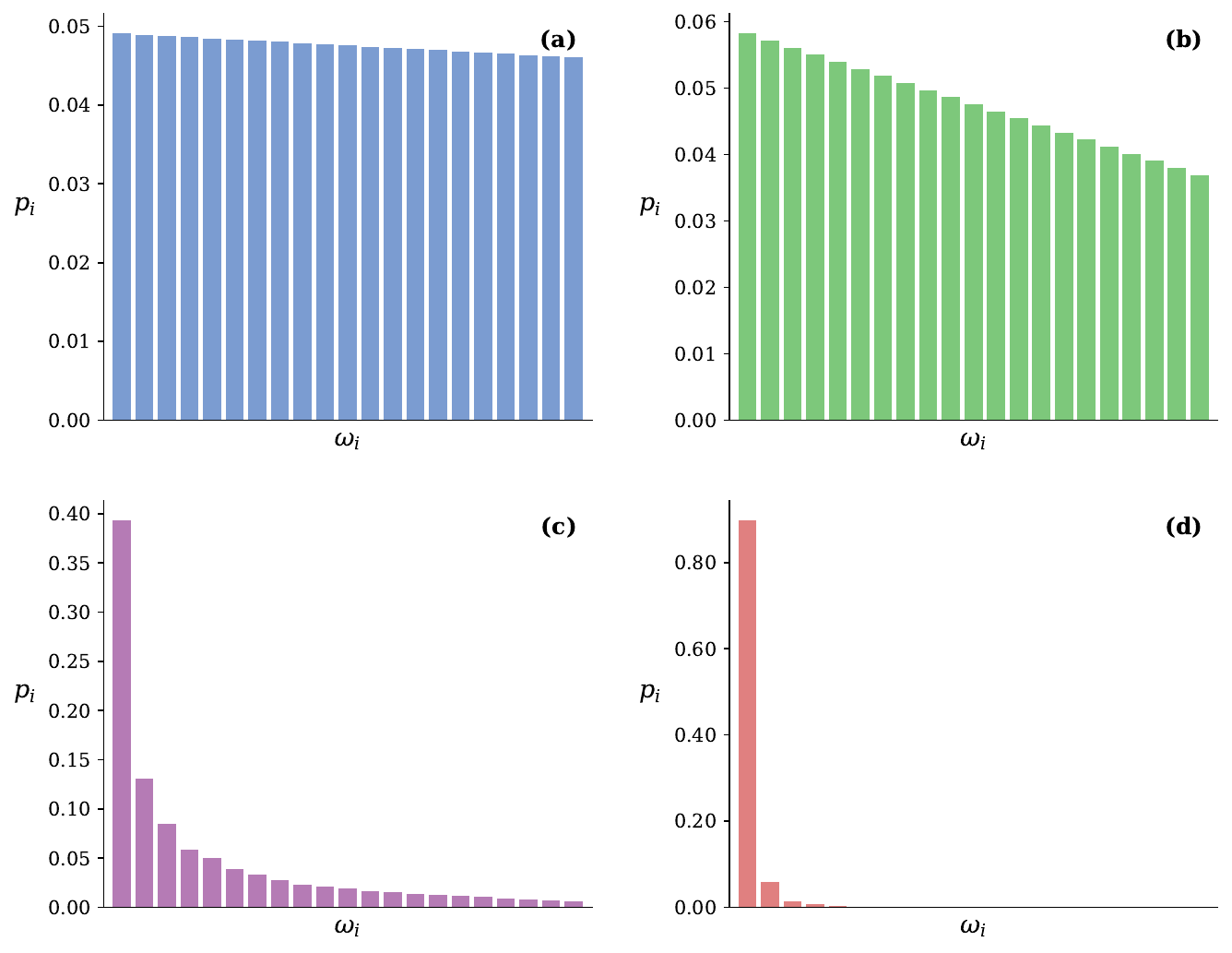}
\caption{\sl {Classical Fröhlich-like condensation.} Normalized energy fractions $p_i$ in the normal modes vs.\ mode frequencies $\omega_i$, for $N=20$ modes. As the energy input rate $S$ increases, deviations from equipartition grow progressively. Panels~(a)--(d) correspond to $S=0.1$ (blue), $1$ (green), $10$ (purple), and $100$ (pink). Equipartition gives equal bar heights; note the increasing concentration of energy in the lowest-frequency mode at large $S$. From Ref.\protect\cite{Nardecchia2018}.}
\label{fig1}
\end{figure}

As $S$ increases, $p_0$ grows while all $p_i$ ($i > 0$) decrease.
The condensation index $E_y(S)$ and the ratio $(p_1/p_0)(S)$ display a
crossover near $S \approx 10$ that sharpens progressively with $N$, with
curves for different $N$ crossing at a common point whose slope grows with
$N$, suggestive of a sharp bifurcation in the large-$N$ limit.
This finite-size sharpening mirrors the behaviour of order-parameter curves
near second-order equilibrium phase transitions.

\begin{figure}[h!]
 \centering
\includegraphics[scale=0.2,keepaspectratio=true,angle=0]{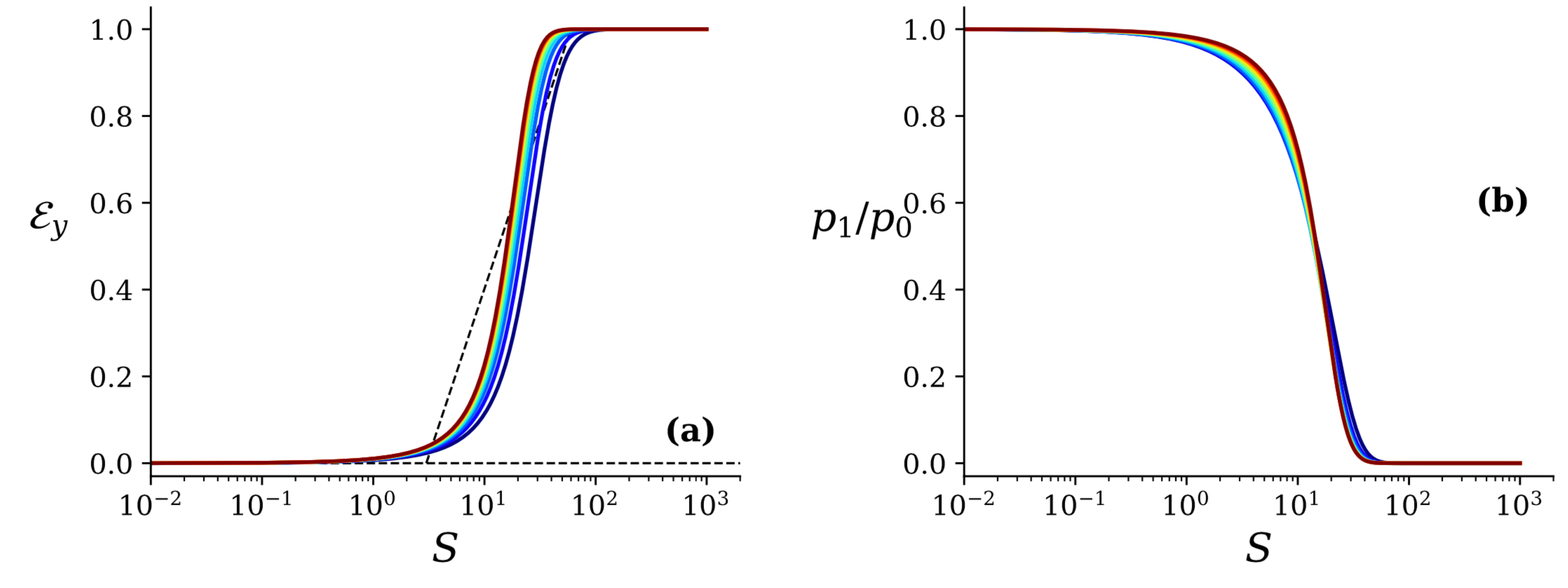}
\caption{\sl {Classical Fröhlich-like condensation.} Condensation index $\mathcal{E}_y$ (left panel) and ratio $p_1/p_0$ (right panel) vs.\ energy input rate $S$, for increasing mode number $N_\mathrm{sys}$: from right to left curves $N= 11, 21, 41, 101, 301$.%[black 11, blue: 21, : 41, green: 101, orange: 201, brown: 301]. 
At equipartition $\mathcal{E}_y=0$, $p_1/p_0=1$; at full condensation $\mathcal{E}_y=1$, $p_1/p_0=0$. The dashed oblique line marks the inflection tangent as a guide to a possible asymptotic bifurcation. From Ref.\protect\cite{Nardecchia2018}. }
\label{fig2}
\end{figure}
The dynamics reveals a hierarchy: the highest-frequency modes lose energy
first at increasing $S$; lower modes follow; eventually all modes deplete
except $\omega_0$, which absorbs the surplus and saturates at $p_0 \to 1$.
The saturation is robust and mode-number independent at high drive.

\subsection{Physical interpretation}

Classical phonon condensation is an instance of a nonequilibrium phase
transition~\cite{Haken1975}: an open system with nonlinear internal couplings,
dissipation, and external drive self-organises into a macroscopically ordered
state when the energy-input rate exceeds a critical value.
The genericity of this mechanism, depending only on nonlinear mode coupling,
a thermal bath, and a source, not on molecular details, suggests the
phenomenon should occur in real macromolecules.
In the condensed state the lowest-frequency mode carries a coherent
oscillation of the entire molecule, activating a giant oscillating dipole
moment that is a necessary prerequisite for long-range electrodynamic
forces~\cite{Preto2015}.

%============================================================
\section{Long-Range Resonant Electrodynamic Interactions}
\label{sec:EDforces}
%============================================================

\subsection{Classical electrodynamic Hamiltonian for two dipoles}

The electrodynamic interaction between two coherently oscillating macromolecules
is modelled at the classical level as a pair of driven harmonic oscillators
coupled through the retarded electromagnetic field of the intervening medium.
Each macromolecule is represented by its dominant polar mode,  an oscillating
electric dipole $\boldsymbol{\mu}_{A,B}(t)$ at natural frequency $\omega_{A,B}$
with effective charge-to-mass ratio $\zeta_{A,B}=Q_{A,B}^2/m_{A,B}$, where
$Q_{A,B}$ and $m_{A,B}$ are the effective charge and mass of each dipole.
This representation is appropriate because the low-frequency collective
vibrations of macromolecules ($\omega/2\pi \sim 0.1$--$1$\,THz) fall well
within the classical regime $k_BT/\hbar\omega \gg 1$ at room temperature,
and the dominant electrodynamic coupling involves the dipole moment of each
molecule rather than higher multipoles.

The coupling is mediated by the electric field that each dipole radiates into
the surrounding medium, here described by the frequency-dependent permittivity
$\varepsilon(\omega)$. The Debye screening that suppresses static electrostatic
forces in the ionic cellular environment becomes ineffective at frequencies
above $\sim\!250$\,MHz~\cite{Preto2015}, so that the oscillating dipole
fields can propagate without ionic screening at the sub-THz frequencies of
interest. In the dipole approximation the equations of motion
are~\cite{Preto2015}
\begin{align}
  \ddot{\boldsymbol{\mu}}_A +
  \omega_A^2\boldsymbol{\mu}_A
  &= \zeta_A \mathbf{E}_B(\mathbf{r}_A,t)\,,
  \label{eq:dipoleA}\\
  \ddot{\boldsymbol{\mu}}_B +
  \omega_B^2\boldsymbol{\mu}_B
  &= \zeta_B \mathbf{E}_A(\mathbf{r}_B,t)\,,
  \label{eq:dipoleB}
\end{align}
where $\mathbf{E}_{A(B)}(\mathbf{r}_{B(A)},t)$ is the electric field created
by dipole $A$ (resp.\ $B$) at the location of its partner. Damping terms
$\gamma_{A,B}\dot{\boldsymbol{\mu}}_{A,B}$ and anharmonic contributions
are present in the full equations of motion~\cite{Preto2015} but are omitted
here as the focus is on the conservative normal-mode structure from which
the interaction potential is derived.
%their role in limiting the range of the interaction is discussed in Sec.~\ref{subsec:damping}.

The field $\mathbf{E}_{A(B)}$ at the position of the partner is linear in the
source dipole and is fully characterised, for a given separation $r$ and
frequency, by the electric susceptibility tensor $\chi_{ij}(r,\omega)$ of the
medium, whose explicit form is obtained from the solution of the
d'Alembert equation for the vector potential in Lorenz gauge~\cite{Preto2015}.
Taking the $z$-axis along the intermolecular separation $\mathbf{r}$, the
tensor is diagonal with elements
\begin{equation}
  \chi'_{11}(r,\omega) = \chi'_{22}(r,\omega)
  = -\frac{\cos(\omega\sqrt{\varepsilon'}r/c)}{\varepsilon'(\omega)r^3}
  \Bigl(1 - \frac{\omega^2\varepsilon'(\omega)r^2}{c^2}\Bigr)
  + \cdots, \quad
  \chi'_{33}(r,\omega) = -2\chi'_{11}(r,\omega)/\bigl(1-\cdots\bigr),
  \label{eq:chi}
\end{equation}
where $\chi'_{ii}$ denotes the real (in-phase) part of $\chi_{ii}$,
which is the component that contributes to the conservative interaction
potential. In the near zone ($r \ll c/\omega$) and for a non-dispersive
background ($\varepsilon' = \text{const}$), Eq.~\eqref{eq:chi} reduces to
the static dipole propagator $\chi'_{ii} \sim \sigma_i/(\varepsilon' r^3)$
(with $\sigma_{1,2}=-1$, $\sigma_3=2$), while at finite frequency
the factor $\varepsilon(\omega)$ renormalises this scaling and retardation
corrections introduce oscillatory $r^{-2}$ and $r^{-1}$ terms whose
importance grows with $r$.

The normal frequencies $\omega_N$ of the dissipation-free coupled system are
obtained by substituting the Fourier ansatz
$\boldsymbol{\mu}_{A,B}(t)=\boldsymbol{\mu}_{A,B}^{(0)}e^{-i\omega_N t}$
into Eqs.~\eqref{eq:dipoleA}--\eqref{eq:dipoleB} and requiring non-trivial
solutions. The resulting secular condition is
\begin{equation}
  \bigl(\omega_A^2 - \omega_N^2\bigr)
  \bigl(\omega_B^2 - \omega_N^2\bigr)
  = \zeta_A\zeta_B\bigl[\chi'_{ii}(r,\omega_N)\bigr]^2,
  \label{eq:normalfreq}
\end{equation}
an implicit equation in $\omega_N$ because the right-hand side itself depends
on $\omega_N$ through the medium response. This implicitness is not a minor
technicality: as shown in Ref.~\cite{Preto2015}, a naive explicit
approximation of the right-hand side at zeroth order in the coupling leads
to the erroneous conclusion, originally reached by Fröhlich,  that
long-range resonant interactions of the form $U\sim 1/r^3$ survive at thermal
equilibrium. The exact treatment via suitable integration shows that these
contributions cancel identically at thermal equilibrium, reducing $U$ to
a short-range $1/r^6$ potential.

The solutions of Eq.~\eqref{eq:normalfreq} are obtained analytically
by a contour-integral method in the complex plane: the Lagrange inversion
theorem, combined with Rouch\'e's theorem, allows one to resolve the
implicit $\omega_N$-dependence of the right-hand side and yields the
normal-mode frequencies as controlled perturbative expansions in the
coupling $\zeta_A\zeta_B[\chi'_{ii}]^2$, uniformly valid across the
resonant ($\omega_A \simeq \omega_B$) and off-resonant
($\omega_A \gg \omega_B$) regimes..
The key result is that the two regimes differ qualitatively in the range of
the resulting interaction: off-resonance the potential scales as $1/r^6$
(short-range, van der Waals-like), whereas at resonance it scales as $1/r^3$
in the near zone and as $1/r$ in the far zone, long-range at all distances.
It is this selectivity of the resonant interaction that makes it a candidate
mechanism for specific biomolecular recognition.
%------------------------------------------------------------------------------------------------------------------
\subsection{Off-resonance and resonant interaction potentials}
The normal-mode frequencies obtained from Eq.~\eqref{eq:normalfreq} determine
the interaction energy through the shift they impart to the bare frequencies
$\omega_{A,B}$.
The qualitative character of this shift depends critically on whether the two
dipoles are in resonance or not.

\paragraph{Off-resonance regime.}
When $\omega_A \gg \omega_B$, the right-hand side of Eq.~\eqref{eq:normalfreq}
is small compared with the unperturbed product
$(\omega_A^2-\omega_B^2)^2$, and the secular equation can be solved
perturbatively.
To leading order the frequency shift of each mode is proportional to
$\zeta_A\zeta_B[\chi'_{ii}]^2 / (\omega_A^2 - \omega_B^2)$, i.e.\ 
\emph{second-order} in the coupling $\chi'_{ii}$.
Since $\chi'_{ii}\sim r^{-3}$ in the near zone, the resulting interaction
potential scales as $U(r) \propto r^{-6}$ 
a short-range, van der Waals--like potential entirely analogous to the
London dispersion interaction between non-resonant fluctuating dipoles.

\paragraph{Resonant regime.}
At resonance ($\omega_A \approx \omega_B = \omega_0$) the structure of
Eq.~\eqref{eq:normalfreq} changes qualitatively.
Setting $\omega_N = \omega_0 + \delta\omega$ and expanding to first order
in $\delta\omega$, the left-hand side becomes $(\omega_0^2-\omega_N^2)^2
\approx 4\omega_0^2(\delta\omega)^2$, so the secular equation reduces to a
perfect square and yields
\begin{equation}
  \omega_{i,\pm}(r) \simeq \omega_0 \pm \sqrt{\zeta_A\zeta_B}\,
  \frac{\chi'_{ii}(r,\omega_0)}{2\omega_0}.
  \label{eq:resshift}
\end{equation}
The frequency shift is now \emph{first-order} in $\chi'_{ii}$, not quadratic.
This is the key distinction from the off-resonance case: the square-root
structure of the resonant secular equation lifts the second-order suppression
and produces a qualitatively stronger coupling.

The total interaction energy is expressed in terms of the adiabatic action
variables $J_{i,\pm} = E_{i,\pm}/\omega_{i,\pm}$ of the two normal modes,
where $E_{i,\pm}$ is the energy stored in the $\pm$ branch of polarisation
component $i$.
Summing over the three spatial components,
\begin{equation}
  U(r) = \sum_{i=1}^3 \Delta\omega_{0,i}(r)\,(J_{i,+} - J_{i,-}).
  \label{eq:Ures}
\end{equation}
In the near zone ($r \ll c/\omega_0$) the susceptibility scales as
$\chi'_{ii} \sim r^{-3}$, so $\Delta\omega_{0,i} \sim r^{-3}$ and
\begin{equation}
  U(r) \;\sim\; \pm\frac{1}{r^3},
\end{equation}
a genuinely \emph{long-range} interaction: the exponent drops from
$-6$ in the off-resonance case to $-3$ at resonance, reflecting the
fact that $U$ is now linear rather than quadratic in the susceptibility
$\chi'_{ii}$.
The sign and magnitude are controlled by the action difference
$J_{i,+} - J_{i,-}$: at thermal equilibrium this difference vanishes
at first order, and the long-range term
disappears, recovering a $1/r^6$ free energy.
Out of equilibrium, when phonon condensation concentrates energy in one
normal mode, $J_{i,+} \gg J_{i,-}$ and the $1/r^3$ interaction is activated.
The explicit behaviour of $\chi'_{ii}(r,\omega)$ as a function of $r$
for representative frequencies in the sub-THz and optical regimes
is shown in Fig.~1 of Ref.~\cite{Preto2015}.

%\subsection{Thermal equilibrium suppresses long-range forces}

A crucial result is that resonant long-range interactions \emph{vanish} at
thermal equilibrium.
At equilibrium the Boltzmann distribution weights the two normal-mode energies
equally, causing the action difference $J_{i,+} - J_{i,-}$ in
Eq.~\eqref{eq:Ures} to vanish at first order, giving a short-range
$1/r^6$ free energy~\cite{Preto2015}.
Long-range resonant electrodynamic forces therefore require an out-of-equilibrium
system, where energy concentration in one normal mode, precisely the
condition delivered by phonon condensation, makes $J_{i,+} \gg J_{i,-}$.

%============================================================
\section{Position-Space Classical Hamiltonian and Molecular Dynamics}
\label{sec:hamMD}
%============================================================
%\subsection{Motivation}
The action-angle Hamiltonian of Sec.~\ref{sec:TDVP} is natural for
perturbative rate equations but remote from $(q,p)$ variables used in
molecular dynamics force fields.
A complementary classical Hamiltonian was formulated directly in
position space~\cite{Preto2024}, opening a path toward connecting
Fr\"ohlich condensation to atomistic MD. This has motivated the study of the following model.
\subsection{The position-space Hamiltonian}
The system Hamiltonian is $H = H_0 + H_{\rm int}$, with free part
\begin{equation}
  H_0 = \sum_{i=1}^N \left(\frac{p_i^2}{2m_i} + \frac{1}{2}m_i\omega_i^2 q_i^2\right)
  + \sum_{k=1}^{N_B} \left(\frac{p_k^{(B)2}}{2m_k^{(B)}}
    + \frac{1}{2}m_k^{(B)}\omega_k^{(B)2} q_k^{(B)2}\right)
  + \sum_{l=1}^{N_S} \left(\frac{p_l^{(S)2}}{2m_l^{(S)}}
    + \frac{1}{2}m_l^{(S)}\omega_l^{(S)2} q_l^{(S)2}\right),
  \label{eq:H0MD}
\end{equation}
and interaction
\begin{equation}
  H_{\rm int}
  = \sum_{ik}\phi_{ik}\,q_i q_k^{(B)}
  + \sum_{il}\xi_{il}\,q_i q_l^{(S)}
  + \sum_{ijk}\lambda_{ijk}\,q_i q_j q_k^{(B)},
  \label{eq:HintMD}
\end{equation}
where $\phi_{ik}$, $\xi_{il}$, $\lambda_{ijk}$ are linear protein-bath,
linear protein-source, and nonlinear protein-protein-bath couplings.
Second-order perturbation theory recovers the Fr\"ohlich rate constants
$\Phi_i$, $\Xi_i$, $\Lambda_{ij}$, all depending on resonance
conditions~\cite{Preto2024}.
%\subsection{Key findings}
Hamiltonian equations are integrated numerically with bath oscillators held
at $T = 300$\,K by a Langevin thermostat~\cite{Preto2024}.
The key findings are:
\textit{i)} Pure Fr\"ohlich resonances ($\omega_i - \omega_j \pm \omega_k^{(B)} = 0$)
produce robust condensation; Lifshits sum resonances
($\omega_i + \omega_j - \omega_k^{(B)} = 0$) destroy it.
\textit{ii)} Replacing the bath-mediated trilinear coupling $\lambda_{ijk}q_iq_jq_k^{(B)}$
by a protein-only coupling $\lambda_{ijk}q_iq_jq_k$ completely suppresses
condensation: bath mediation is physically essential, not a technical device.
\textit{iii)} Strong condensates are found at $T = 300$\,K, contradicting an earlier
study~\cite{Reimers2009} that used constant coupling coefficients.
The discrepancy traces to the coupling parametrisation: coefficients
proportional to $\sqrt{m_i m_k^{(B)} \omega_i \omega_k^{(B)}}$ ensure all
modes contribute equally to the interaction energy at equipartition.

\begin{figure}[h!]
 \centering
\includegraphics[scale=0.5,keepaspectratio=true,angle=0]{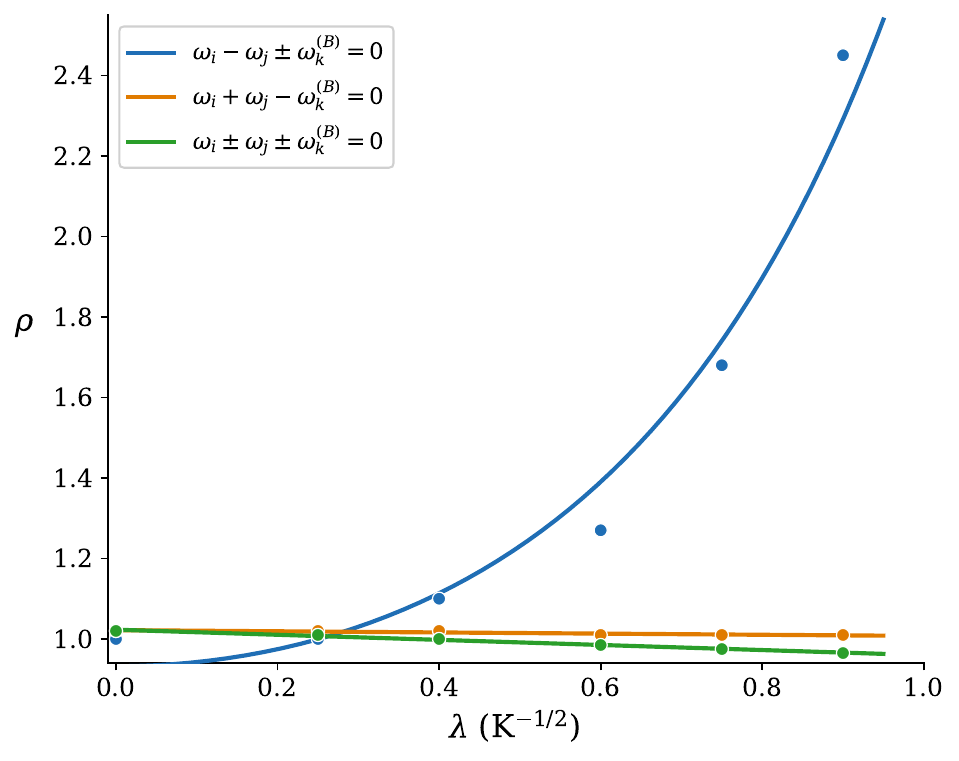}
\caption{\sl  Condensation index $\rho$ of a Fröhlich system from Hamiltonian dynamics, for coupling coefficients $\lambda_{ijk}$  under three resonance conditions: $\omega_i - \omega_j \pm \omega_k^{(B)} = 0$ (Fröhlich), $\omega_i + \omega_j - \omega_k^{(B)} = 0$ (Lifshits), and $\omega_i \pm \omega_j \pm \omega_k^{(B)} = 0$ (combined).  From Ref.\protect\cite{Preto2024}.}
\label{fig4}
\end{figure}
\begin{figure}[h!]
 \centering
\includegraphics[scale=0.105,keepaspectratio=true,angle=0]{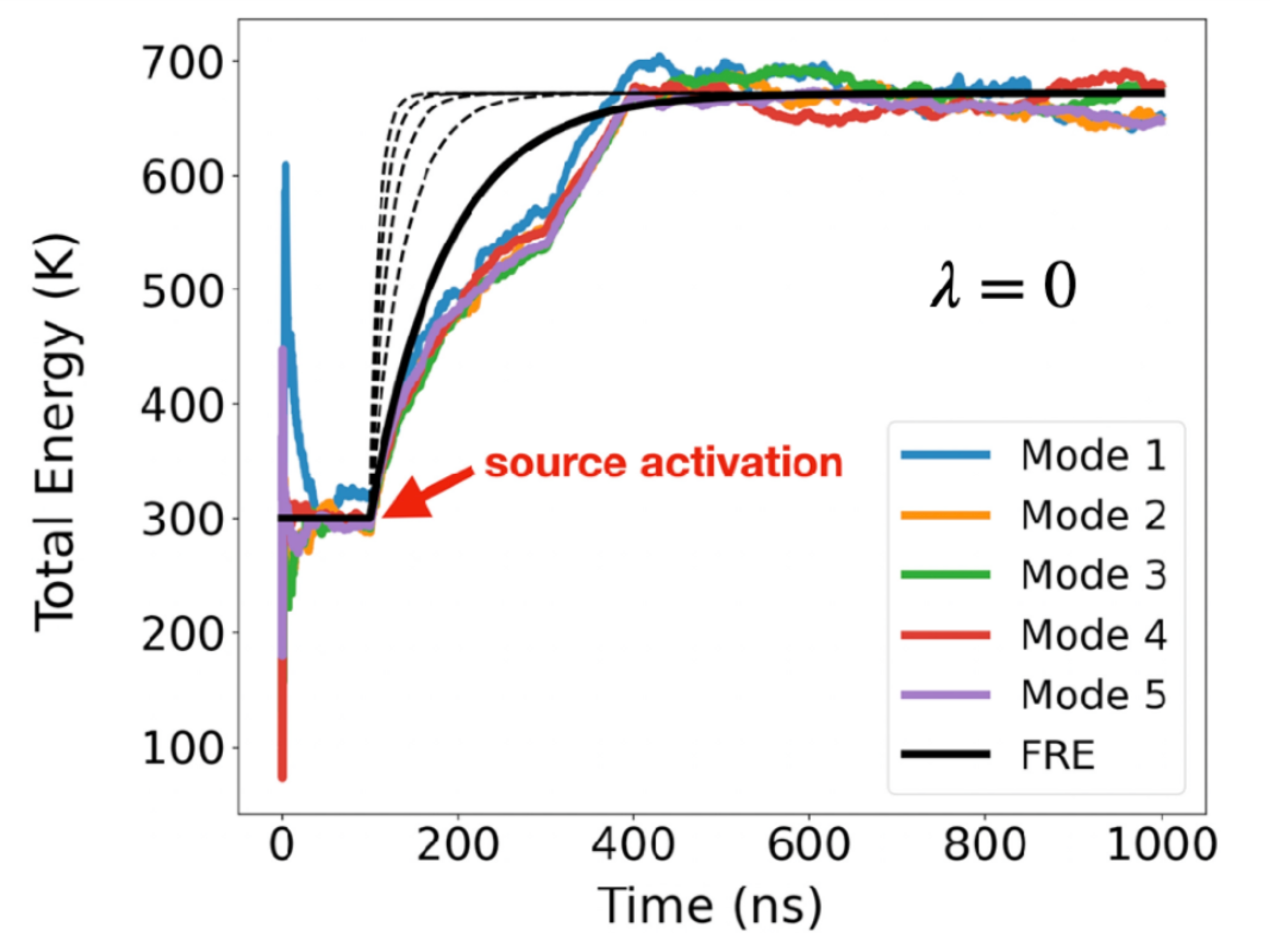} 
\includegraphics[scale=0.1045,keepaspectratio=true,angle=0]{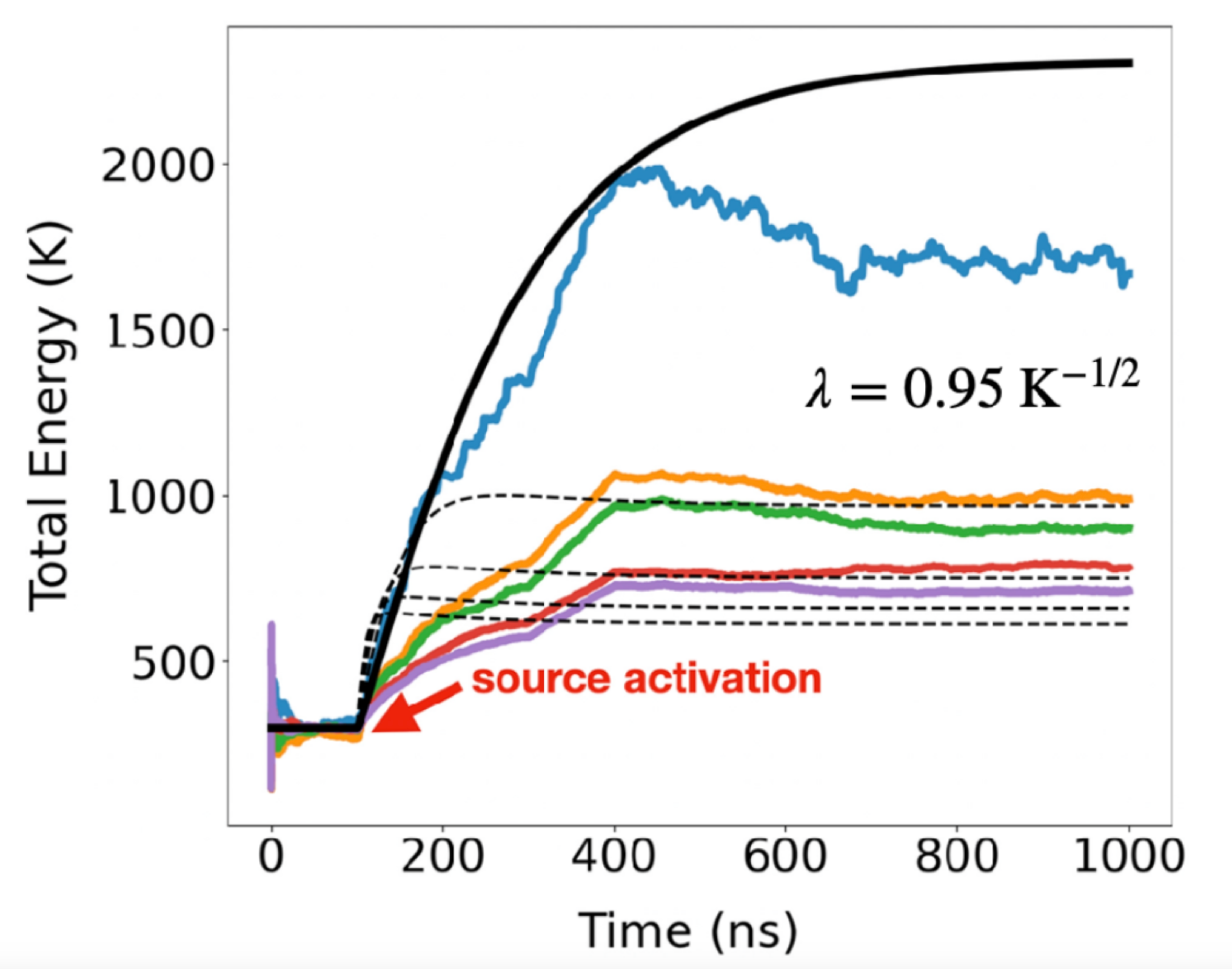}
\caption{\sl  Time evolution of total energies (kinetic + potential) in a Fröhlich system of nine protein modes (first five shown), computed as 300~ns moving averages from Hamiltonian dynamics. Parameters: bath $T=300$~K, source $T_S=3000$~K, protein frequencies $\omega_1=0.2$~THz to $\omega_9=1$~THz in 0.1~THz steps, $\phi_{ik}$, $\xi_{il}$, $\lambda_{ijk}$ from Eq.~(5) of Ref.\protect\cite{Preto2024} with $\phi=1.0$, $\xi=0.4$ (from 100~ns), unit masses, thermostat friction $0.1$~ps$^{-1}$. Black curves: Fr\"ohlich rate equations predictions from Eq.~(\ref{eq:rate}) with $\alpha=0.02$~ps (solid: mode~1; dashed: higher modes). Left panel: $\lambda=0$; right panel: $\lambda=0.95$~K$^{-1/2}$.  From Ref.\protect\cite{Preto2024}.}
\label{fig5}
\end{figure}

%============================================================
\section{Electron-Phonon Hamiltonian and DNA--Protein Co-Resonance}
\label{sec:DNA}
%============================================================

\subsection{Motivation and biological context}

Beyond collective vibrational (phonon) degrees of freedom, the electronic
degrees of freedom of biomolecules provide a further and conceptually
distinct channel through which selective long-range electrodynamic
interactions can be generated~\cite{Faraji2025}.
This possibility is motivated by three converging lines of evidence.
First, the Resonant Recognition Model (RRM)~\cite{Cosic1994,Veljkovic1985}
showed empirically that a Fourier analysis of the electron-ion interaction
potentials (EIIP) along a protein or DNA sequence produces spectral peaks
that correlate with biological activity.
Second, electron transport along DNA and along protein backbones is an
experimentally confirmed phenomenon~\cite{Giese2000,GrayWinkler1996}.
Third, a previous study~\cite{Faraji2021} established that the electron
current flowing along a DNA fragment under external energy supply can
exhibit either a broad or a sharply peaked frequency spectrum depending on
the excitation site and energy, suggesting that co-resonance between DNA
and enzyme currents could provide a sequence-specific electrodynamic signal.

\subsection{The Davydov--Holstein--Fr\"ohlich Hamiltonian}

To model electron--phonon motion along a biomolecular chain of $N$ sites
(nucleotides or amino acids), the following Hamiltonian is
adopted~\cite{Faraji2025}:
\begin{equation}
  \hat{H} = \hat{H}_{\rm el} + \hat{H}_{\rm ph} + \hat{H}_{\rm int},
  \label{eq:HDavydov}
\end{equation}
with electronic part
\begin{equation}
  \hat{H}_{\rm el} = \sum_{n=1}^{N}
     \Bigl[ E_0\hat{B}^\dagger_n\hat{B}_n
       + \epsilon\langle\hat{B}^\dagger_n\hat{B}_n\rangle\hat{B}^\dagger_n\hat{B}_n
       + J_n\bigl(\hat{B}^\dagger_n\hat{B}_{n+1} + \hat{B}^\dagger_n\hat{B}_{n-1}\bigr)
     \Bigr],
  \label{eq:Hel}
\end{equation}
phononic part (with anharmonic term)
\begin{equation}
  \hat{H}_{\rm ph}
  = \frac{1}{2}\sum_n
    \left[
      \frac{\hat{p}_n^2}{M_n}
      + \Omega_n(\hat{u}_{n+1} - \hat{u}_n)^2
      + \frac{1}{2}\mu(\hat{u}_{n+1} - \hat{u}_n)^4
    \right],
  \label{eq:Hph}
\end{equation}
and electron-phonon interaction
\begin{equation}
  \hat{H}_{\rm int}
  = \sum_n \chi_n(\hat{u}_{n+1} - \hat{u}_n)\hat{B}^\dagger_n\hat{B}_n.
  \label{eq:Hint}
\end{equation}
Here $\hat{B}_n$, $\hat{B}^\dagger_n$ are lowering and raising operators
at lattice site $n$; $E_0$ is the initial electron excitation energy;
$\epsilon$ is the nonlinear electron self-trapping constant; $J_n$ is the
site-dependent electron tunnelling amplitude; $\hat{u}_n$, $\hat{p}_n$ are
the longitudinal displacement and momentum of the $n$-th nucleotide or amino acid,
according to the kind of molecule, DNA or protein, respectively; $\Omega_n$
is the site-dependent spring constant; $M_n$ is the site mass; $\mu$ is the
paqrticle-particle (nucleotide or amino acid) anharmonic coupling; and $\chi_n$ is the site-dependent
electron-phonon coupling.
The coupling parameters $J_n$ and $\chi_n$ are site-dependent, encoding 
the specific sequence of nucleotides (for DNA) or amino acids (for protein).

\subsection{TDVP applied to the Davydov ansatz}

The TDVP is again the instrument of choice.
The trial wave function is written in the Davydov factorised form
\begin{equation}
  |\psi(t)\rangle = |\Psi(t)\rangle|\Phi(t)\rangle,
  \label{eq:Davansatz}
\end{equation}
where the electronic part $|\Psi(t)\rangle$ is a single-excitation state
\begin{equation}
  |\Psi(t)\rangle = \sum_n C_n(t)\hat{B}^\dagger_n|0\rangle_{\rm el},
  \label{eq:Psielectron}
\end{equation}
with probability amplitudes $C_n(t)$, and the phononic part $|\Phi(t)\rangle$
is a displaced coherent state
\begin{equation}
  |\Phi(t)\rangle = e^{-\frac{i}{\hbar}\sum_n[\beta_n(t)\hat{p}_n - \pi_n(t)\hat{u}_n]}
  |0\rangle_{\rm ph},
  \label{eq:Phiphonon}
\end{equation}
so that $\langle\Phi|\hat{u}_n|\Phi\rangle = \beta_n(t)$ and
$\langle\Phi|\hat{p}_n|\Phi\rangle = \pi_n(t)$.

Applying the stationary-action condition $\delta S = 0$ with the Lagrangian
$L(t) = i\hbar\langle\psi|\partial_t|\psi\rangle - \langle\psi|\hat{H}|\psi\rangle$
yields equations of motion for the variational parameters
$(C_n, C_n^*, \beta_n, \pi_n)$:
\begin{equation}
  i\hbar\dot{C}_n = \partial_{C_n^*} H, \qquad
  \dot{\beta}_n = \partial_{\pi_n} H, \qquad
  \dot{\pi}_n = -\partial_{\beta_n} H,
  \label{eq:TDVP_eqs}
\end{equation}
where $H = \langle\psi|\hat{H}|\psi\rangle$ is the classical Hamiltonian
obtained as the expectation value.

Evaluating $H$ explicitly~\cite{Faraji2025}:
\begin{align}
  H &= \sum_n
       \Bigl[
         E_0|C_n|^2 + \epsilon|C_n|^4
         + J_n(C_n^*C_{n+1} + C_{n+1}^*C_n)
  \nonumber\\
  &\qquad\quad
         + \frac{1}{2}\!\left(
             \frac{\pi_n^2}{M_n}
             + \Omega_n(\beta_{n+1} - \beta_n)^2
             + \frac{\mu}{2}(\beta_{n+1}-\beta_n)^4
           \right)
  \nonumber\\
  &\qquad\quad
         + \chi_n(\beta_{n+1} - \beta_n)|C_n|^2
       \Bigr].
  \label{eq:Hclass_davydov}
\end{align}
The equations of motion that follow from Eqs.~\eqref{eq:TDVP_eqs} and
\eqref{eq:Hclass_davydov} are formally classical but describe the time
evolution of quantum expectation values:
\begin{align}
  i\hbar\dot{C}_n
  &= \bigl[E_0 + 2\epsilon|C_n|^2 + \chi_n(\beta_{n+1} - \beta_n)\bigr]C_n
  \nonumber\\
  &\quad
     + J_nC_{n+1} + J_{n-1}C_{n-1},
  \label{eq:Cdot}
\end{align}
\begin{align}
  M_n\ddot{\beta}_n
  &= \Omega_n\beta_{n+1} + \Omega_{n-1}\beta_{n-1}
     - \Omega_{n-1}\beta_n - \Omega_n\beta_n
  \nonumber\\
  &\quad
     + \chi_n|C_n|^2 - \chi_{n-1}|C_{n-1}|^2
  \nonumber\\
  &\quad
     + \mu\bigl[(\beta_{n+1} - \beta_n)^3 - (\beta_n - \beta_{n-1})^3\bigr].
  \label{eq:betaddot}
\end{align}
Equations~\eqref{eq:Cdot}--\eqref{eq:betaddot} couple the quantum
probability amplitudes for electron propagation to the classical phonon
displacements of the chain, encoding the full electron--phonon dynamics
of the specific biomolecular sequence.

\subsection{Physical parameters and numerical strategy}

In Ref.~\cite{Faraji2025} the DNA--protein system studied is a 66~bp DNA oligonucleotide interacting
with the EcoRI restriction enzyme (276 amino acids).
The site-dependent parameters $J_n$ and $\chi_n$ are determined from the
electron-ion interaction potentials (EIIP) of each nucleotide
(A, T, G, C) and each amino acid, tabulated in Refs.~\cite{Cosic1994,Veljkovic1985,CosicBook}.
Tunnelling amplitudes are estimated from the
transmission probability $P(n \to n \pm 1)$ through the potential barrier
between adjacent sites; the site-dependent electron-phonon coupling is
estimated as $\chi_n = dE/dx = (E_{n+1} - E_n)/a$, where $a$ is the
nearest-neighbour distance between particles ($a = 3.4$\,\AA\ for nucleotides of DNA, $4.5$\,\AA\ for the
amino acids of enzyme).

The particle displacements are initialised at thermal equilibrium at
$T = 310$\,K:
\begin{equation}
  \langle|b_n(0)|\rangle_n = \sqrt{\frac{k_BT}{\hbar\omega\Omega'}},
  \qquad
  \langle|\pi_n(0)|\rangle_n = \sqrt{\frac{k_BT}{\hbar\omega}}.
  \label{eq:thermalinit}
\end{equation}
The electron wavefunction is initialised as a localised sech pulse
centred at site $n_0$.
Integration uses a symplectic leapfrog scheme for the phonon variables
combined with an Euler predictor-corrector for the electron amplitudes,
with relative energy conservation $\Delta E/E \sim 10^{-6}$.

\subsection{Cross-spectral co-resonance and sequence specificity}

The primary observable is the cross Fourier spectrum of the electron
current densities along the two chains.
The average electron current along chain $\alpha$ ($\alpha = 1$ for DNA,
$\alpha = 2$ for EcoRI) is
\begin{equation}
  i_\alpha(t)
  = \frac{e\hbar}{2N_\alpha a_\alpha m_e i}
    \sum_{j=1}^{N_\alpha}
    \left(
      \Psi^*_\alpha(x_j,t)\,\frac{\Psi_\alpha(x_{j+1},t) - \Psi_\alpha(x_{j-1},t)}{2}
      - \text{c.c.}
    \right),
  \label{eq:current}
\end{equation}
and the cross-spectrum $\tilde{i}_1^*(\nu)\tilde{i}_2(\nu)$ is computed from
their Fourier transforms.

The central finding~\cite{Faraji2025} is illustrated schematically as follows.
When the EcoRI recognition site on the DNA is the canonical palindromic
sequence $3'$-CTTAA$|$G-$5'$ (where $|$ marks the cleavage site), the
cross-spectrum displays a \emph{sharp co-resonance peak} at
$\nu \approx 20$--$29$\,THz (depending on initial conditions), in
quantitative agreement with the RRM prediction~\cite{Cosic1994}.

\begin{figure}[h!]
\includegraphics[width=0.3\columnwidth]{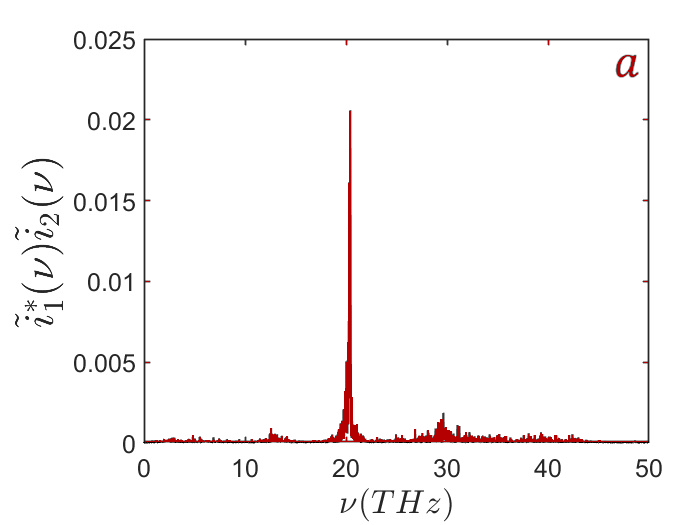}
\includegraphics[width=0.3\columnwidth]{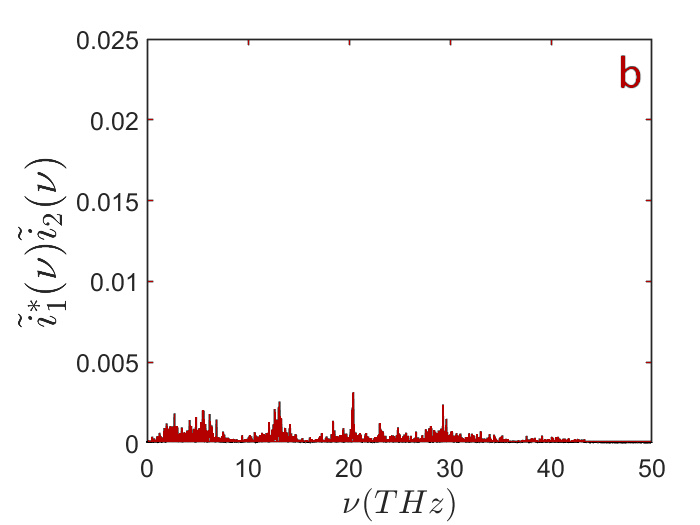}
\includegraphics[width=0.3\columnwidth]{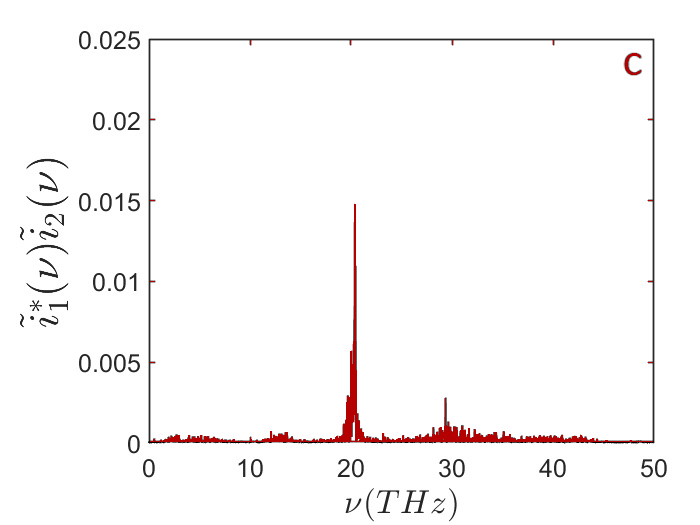}
\includegraphics[width=0.3\columnwidth]{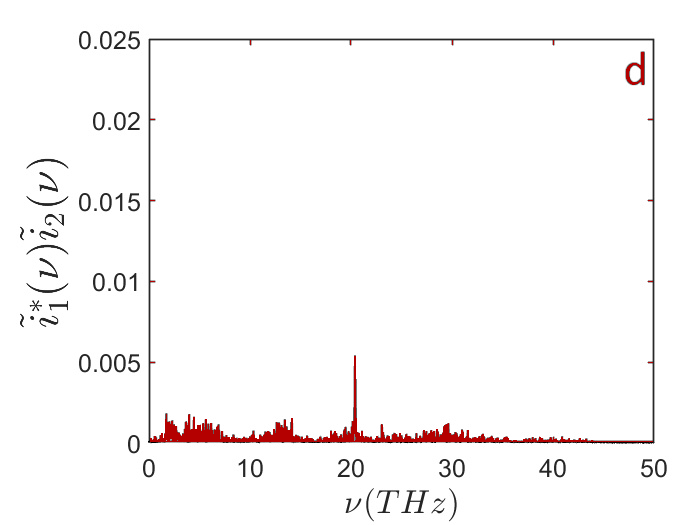} 
   \caption{\sl Cross frequency spectra between a DNA strand with $N_{1}=66$ nucleotides and the \textit{Eco}RI enzyme with $N_{2}=276$ amino acids.  Panels show DNA strand containing a) the canonical target CTTAAG recognition site, b) cyclic permutation of restriction site in a) to AGCTTA, c) one-nucleotide change from a) with its base-pair complement to CATAAG, and d) two-nucleotide change from a) with their base-pair complements to GTTAAC. From Ref.\protect\cite{Faraji2025}.}
 \label{37823725}
 \end{figure}

When the recognition sequence is randomised (e.g., AGCTTA, TCATGA, AGATCT),
the sharp peak disappears and is replaced by a broad, noisy spectrum.
When one nucleotide of the recognition site is exchanged with its
complement (e.g., CATAAG), the peak undergoes a modest broadening; two
exchanges (e.g., GTTAAC) broaden it more.
The co-resonance is therefore robustly and specifically associated with
the canonical recognition sequence.

 \begin{figure}[h!]
\includegraphics[width=0.3\columnwidth]{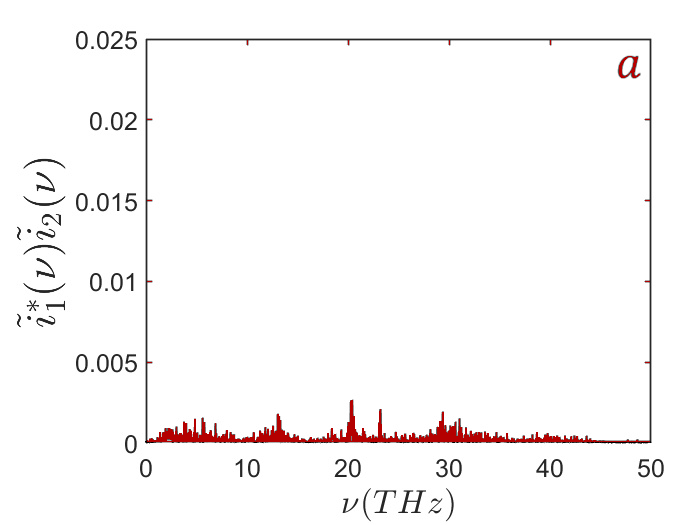}   
\includegraphics[width=0.3\columnwidth]{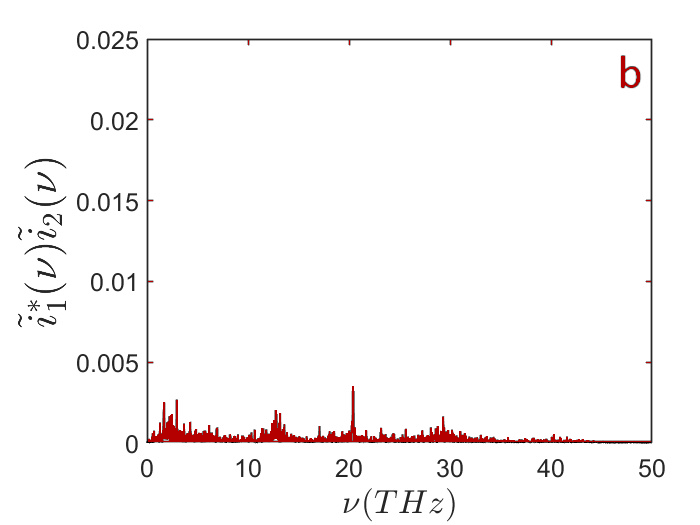}  
\includegraphics[width=0.3\columnwidth]{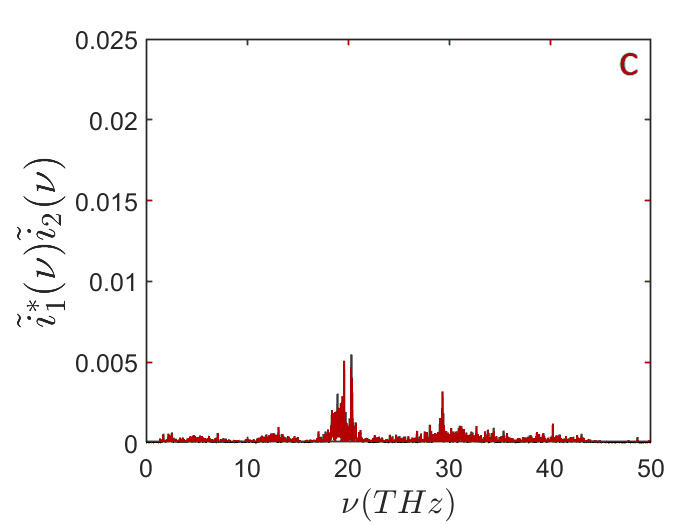} 
   \caption{\sl Cross frequency spectra between a DNA strand and the \textit{Eco}RI enzyme, under the same substrate and initial conditions of Fig.~\ref{37823725}. Panels show DNA strand containing:  a) randomized restriction site TCATGA;  b) one-nucleotide change from the canonical target CTTAAG to CTTAAC; c) two-nucleotide change from the canonical target CTTAAG to CATATG. From Ref.\protect\cite{Faraji2025}.}
 \label{37823725bis}
 \end{figure}
 
   \begin{figure}[h!]
  \includegraphics[width=0.34\columnwidth]{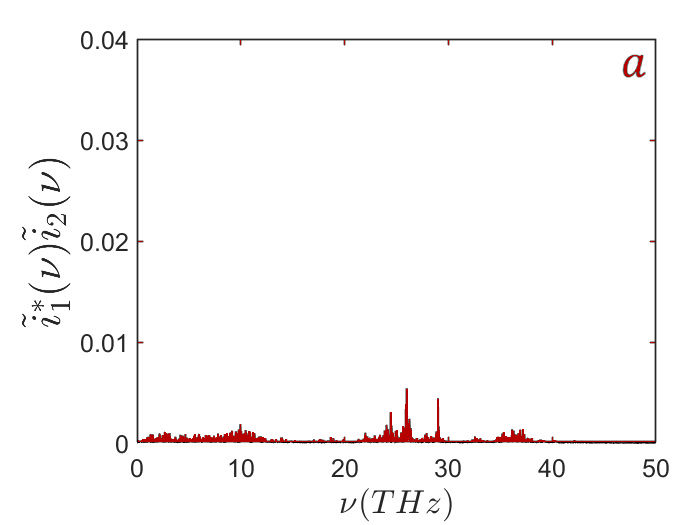}  
     \includegraphics[width=0.34\columnwidth]{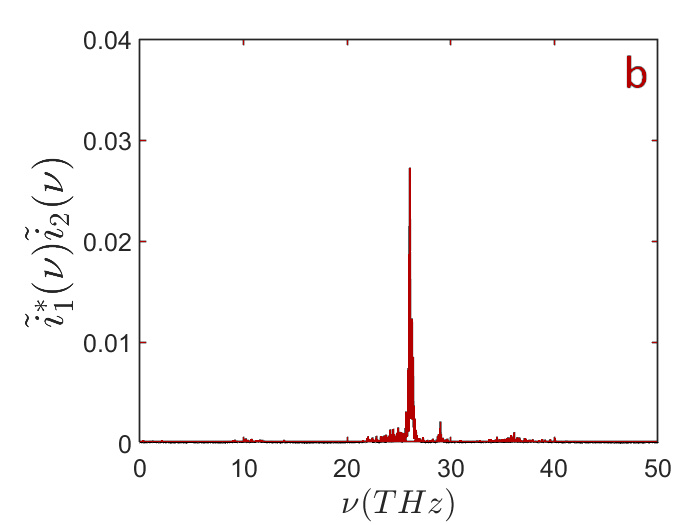}   
       \includegraphics[width=0.34\columnwidth]{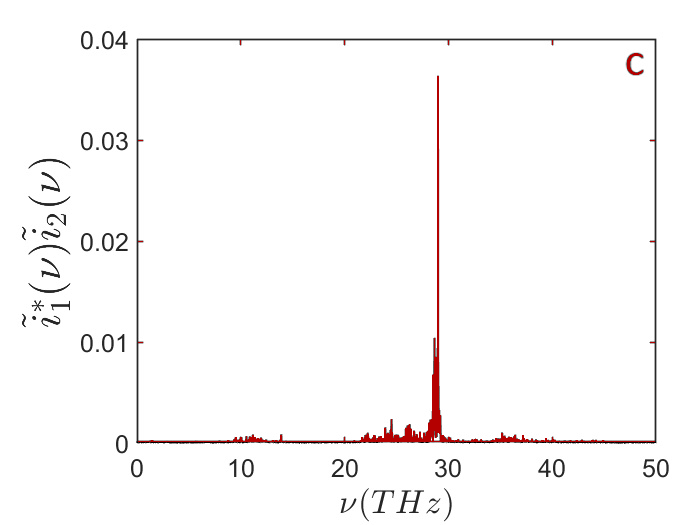}      
    \caption{\sl Cross frequency spectra between a DNA strand and the \textit{Eco}RI enzyme, under the same substrate and initial conditions of Fig.~\ref{37823725}. Panels show DNA strand containing a e) randomized restriction site AGATCT, f) one-nucleotide change from the canonical target CTTAAG to CATAAG, g) two-nucleotide change from the canonical target CTTAAG to GTTATG. From Ref.\protect\cite{Faraji2025}.}
  \label{37163715bis}
  \end{figure}
 
The model also discriminates between point-mutation variants with known
biological activity~\cite{Faraji2025}: promiscuous mutations
(Ala$^{138}\to$Thr, Glu$^{192}\to$Lys, His$^{114}\to$Tyr) that allow
binding to near-cognate sequences still yield a sharp co-resonance peak,
while the catalytically inactivating mutation Asp$^{91}\to$Asn produces
a sizeable decrease of the peak.
These findings are coherent with the experimental biochemical data.

  \begin{figure}[h!]
  \includegraphics[width=0.34\columnwidth]{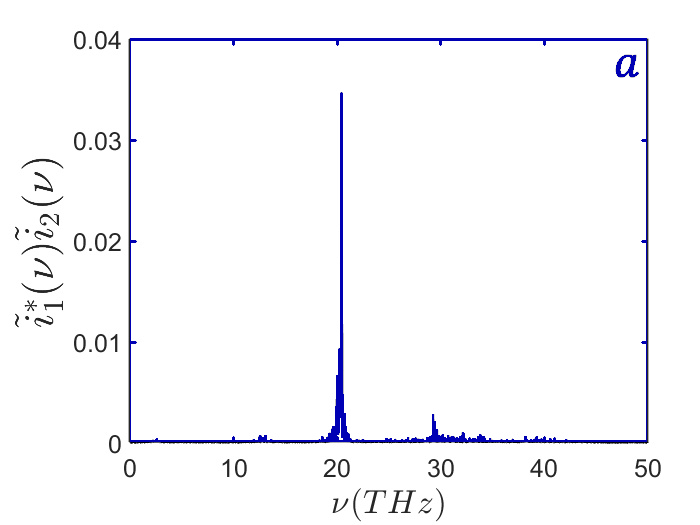}  
     \includegraphics[width=0.34\columnwidth]{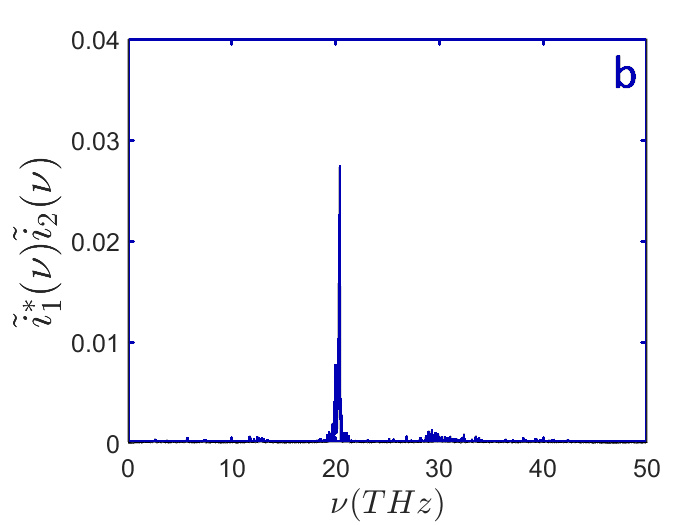}   
       \includegraphics[width=0.34\columnwidth]{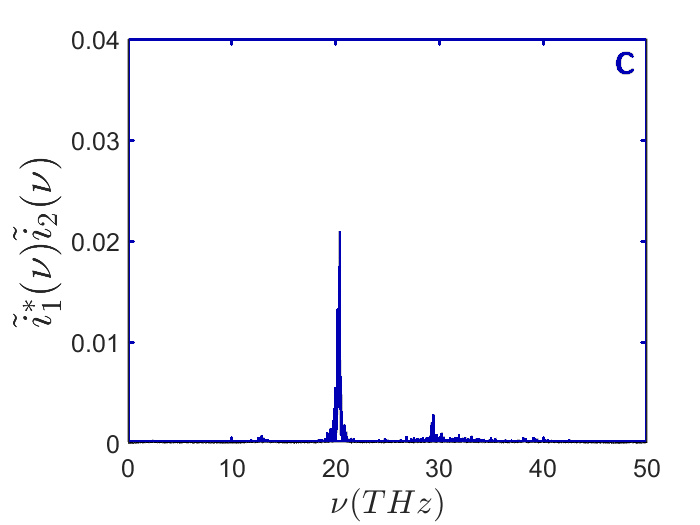}   
       \includegraphics[width=0.34\columnwidth]{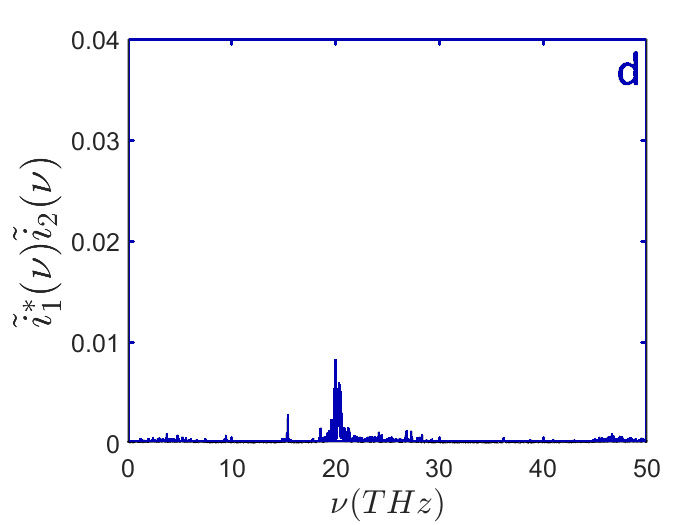}
\caption{ {\sl Cross frequency spectra between the original substrate and the \textit{Eco}RI mutants with the single point mutations;  a) $Ala^{138}\to Thr$; b) $Glu^{192}\to Lys$; c) $His^{114}\to Tyr$; and d) $Asp^{91}\to Asn$. The initial conditions are the same of Figure~\ref{37823725}. From Ref.\protect\cite{Faraji2025}.}}
  \label{fig:mutations}
  \end{figure}
  
\subsection{Physical interpretation and comparison with the vibrational channel}

The TDVP-derived equations of motion~\eqref{eq:Cdot}--\eqref{eq:betaddot}
are formally classical but arise directly from a quantum Hamiltonian via the
same variational dequantization procedure applied in Sec.~\ref{sec:TDVP} to
the Wu--Austin model.
The two applications of the TDVP operate on different physical degrees of
freedom (collective vibrational modes vs.\ electron-phonon propagation) and
at different energy scales (sub-THz vs.\ tens of THz), but share the same
mathematical structure: a product ansatz over coherent (or single-excitation)
states, a classical Lagrangian obtained as the expectation value of the
Hamiltonian, and equations of motion for the classical parameters governing
the trial state.

The co-resonance reported here constitutes a distinct and complementary
mechanism for selective electrodynamic interaction between biomolecules.
In the vibrational channel (Secs.~\ref{sec:condensation}--\ref{sec:EDforces}),
the relevant quantity is the oscillating dipole moment carried by a collective
phonon mode driven into a condensed state by external energy input.
In the electronic channel (this section), the relevant quantity is the
oscillating electron current density whose spectral overlap with that of the
partner molecule determines the strength of the co-resonant coupling.

Both channels require out-of-equilibrium conditions for their activation:
the phonon channel requires energy input exceeding the condensation threshold;
the electronic channel requires electron excitation, provided in the cell,
for example, by redox reactions, ATP-driven ion currents, or biophoton
emission~\cite{Faraji2025}.
It is plausible that both channels are simultaneously active in vivo,
and that they reinforce each other in determining the selectivity and range
of biomolecular recognition.

The mediating role of the aqueous environment is highlighted by the
Jaynes--Cummings-like interaction Hamiltonian~\cite{Faraji2025}
%\begin{equation}
 $ H_{\rm int} = \hbar\sqrt{N}\,\gamma(a^\dagger S^- + a S^+)$,
 % \label{eq:JC}
%\end{equation}
where $a^\dagger$, $a$ are the creation and annihilation operators for the
DNA radiative electric field, $S^\pm$ are raising/lowering operators for
the collective orientational state of the $N$ surrounding water dipoles,
and the coupling scales as $\sqrt{N}$, suggesting that the large number of
water molecules in a physiological domain provides a protective gap against
thermalization for the long-range correlations~\cite{Kurian2018,Faraji2025}.

%============================================================
\section{Experimental Evidence}
\label{sec:experiment}
%============================================================

\subsection{Experimental strategy}

The unprecedented experimental programme outlined in the present section
has been focused on two observables predicted by theory.
First, an out-of-equilibrium macromolecule undergoing phonon condensation
should display a sharp sub-THz absorption feature absent at thermal
equilibrium, with a threshold in the energy-input rate~\cite{Nardecchia2018}.
Second, if two such macromolecules interact via resonant electrodynamic
forces, the collective oscillation frequency of each should shift as
$\Delta\nu \propto 1/\langle r\rangle^3$~\cite{Preto2015,Lechelon2022},
and above a critical concentration a clustering phase transition should
occur~\cite{Lechelon2022}.
\begin{figure}[h!]
%\centering
\includegraphics[scale=0.17,keepaspectratio=true,angle=0]{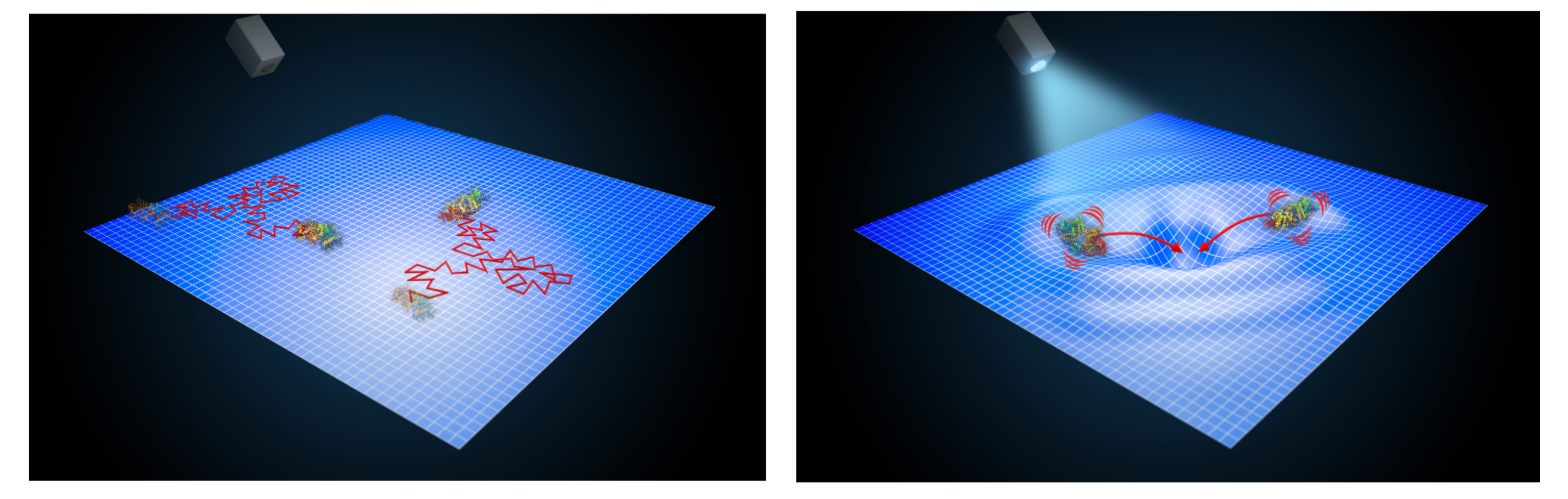}
\caption{{\sl Long-range electrodynamic interactions - Principle and experimental approaches}  At thermal equilibrium, macromolecules show a Brownian diffusive motion in solution (left panel). By switching-on an external energy source, molecules are in an out-of-thermal equilibrium collective vibrational state that can generate ED forces through associated large dipolar resonant oscillations (right panel). From Ref.\protect\cite{Lechelon2022}. }
\label{prot-prot-interaction}
\end{figure}

Out-of-equilibrium conditions are achieved by optical pumping of
fluorochrome-labelled proteins with a continuous-wave Argon laser (488\,nm),
inducing ``proteinquake'' energy transfer to the vibrational modes.
Two complementary THz near-field spectroscopy setups, a microwire probe
and a plasmonic rectenna, operated in two independent laboratories, 
together with fluorescence correlation spectroscopy (FCS) serve as detectors.

\subsection{Phonon condensation in BSA}

The first experimental test of the classical phonon condensation scenario
was performed with bovine serum albumin (BSA), a globular protein of
$67$\,kDa chosen as a well-characterised model system~\cite{Nardecchia2018}.
Because BSA has no endogenous chromophore suitable for efficient optical
pumping, five to six Alexa\,488 fluorochrome molecules were covalently
attached to its lysine residues and excited by an argon-ion laser at
$488$\,nm.
The energy difference between absorbed and re-emitted photons
($\approx 0.19$\,eV per fluorochrome per incident photon) provides the
energy input that drives the protein out of thermal equilibrium via the
proteinquake mechanism~\cite{Ansari1985}.

THz near-field absorption spectroscopy was performed simultaneously in two
independent laboratories using complementary probe technologies: a
$12\,\mu$m-diameter microcoaxial wire (Montpellier) and a plasmonic
bow-tie rectenna coupled to a plasma-wave FET (Rome).
Both setups operate in the $0.22$--$0.33$\,THz range, with a spectral
resolution below $300$\,Hz from continuous-wave Virginia\,Diodes sources.
A principal resonance at $\nu = 0.314$\,THz was reproducibly observed in
both laboratories only when (i)~Alexa\,488 was bound to BSA and
(ii)~the laser was switched on; control measurements on labelled BSA
without laser illumination, on free dye without protein, and on
unlabelled BSA with illumination all showed no spectral feature.

The resonance frequency is in quantitative agreement with the lowest
spheroidal deformation mode of an elastic sphere,
\begin{equation}
  \nu_0 = \frac{1}{2\pi}
  \left[\frac{2(2l+1)(l-1)E}{\rho R_H^2}\right]^{1/2},
  \quad l=2,
  \label{eq:spheroid}
\end{equation}
giving $\nu_0 \approx 0.308$\,THz using the independently measured
Young modulus $E = 6.75$\,GPa and hydrodynamic radius
$R_H = 35$\,\AA~\cite{Nardecchia2018}.

A complementary, rough estimate treats the BSA as a sphere splitted into two
parts of masses $m = 33$\,kDa joined by a spring of stiffness $k = EA_0/\ell_0$,
giving $\nu \approx 0.300$\,THz~\cite{Nardecchia2018}, within 5\% of
the observed resonance, providing independent confirmation that the
feature is a global deformation mode of the whole molecule.
The two weaker resonances observed at $0.278$ and $0.285$\,THz can be
tentatively assigned to torsional modes at frequencies
$\nu_t = \nu_0[(2l+3)/(2(2l+1))]^{1/2}$ for $l=2$ and $l=3$,
which yield $0.257$ and $0.246$\,THz respectively; the remaining
discrepancy is consistent with the non-spherical shape of BSA
introducing different moments of inertia along different axes.
The absorption line profile is well described by the Lorentzian expected
for a damped harmonic oscillator (Eq.~52 of Ref.~\cite{Nardecchia2018}),
with a quality factor $Q = \Delta\nu/\nu \approx 50$.

The threshold and saturation behaviour of the resonance intensity as
a function of laser power matches the theoretical prediction from the
nonlinear rate equations~\eqref{eq:ratend}.

A consistency estimate of the onset timescale (several minutes of
illumination before the resonance reaches its final amplitude) can be
obtained from the energy balance
\begin{equation}
  \frac{dE}{dt} = -\frac{2}{3}\frac{(Ze)^2|\ddot{x}|^2}{c^3}
                  - \Gamma + W ,
  \label{eq:Ebalance}
\end{equation}
where $W$ is the optical energy input rate and the first term on the
right is the bremsstrahlung (radiative) loss.
With five to six Alexa\,488 molecules per BSA each receiving roughly
$120$--$300$ photons per second at $500\,\mu$W of laser power,
the upper bound for $W$ is $3.8$--$9.5\times 10^{-11}$\,erg\,s$^{-1}$,
almost coinciding with the estimated radiative loss of
$3.25\times 10^{-12}$--$1.7\times 10^{-11}$\,erg\,s$^{-1}$~\cite{Nardecchia2018}.
Because the two rates are so nearly balanced, energy accumulates slowly
in each molecule, and several minutes elapse before the condensation
threshold is crossed.
Once the condensed phase is established, the collective low-frequency
oscillation carries a very large effective dipole moment: requiring
the bremsstrahlung losses of the condensed oscillation to be balanced
by $W$ implies an oscillating dipole in the range $14{,}500$--$23{,}000$\,Debye,
corresponding to an effective charge $Z \approx 290$--$460$ elementary
charges~\cite{Nardecchia2018}.  This is fully consistent with the
strong THz absorption feature emerging against the water background
($\approx 2000$\,dB/cm extinction).

Finally, we note the biological relevance of these estimates.
In living cells, a protein molecule undergoes roughly $10^6$ collisions
per second with ATP molecules at a standard intracellular concentration
of $\sim 1$\,mM.
If even 1--2\% of those collisions transfer the $\approx 50$\,kJ/mol
of ATP hydrolysis free energy to the protein, the available power is
$\sim 8\times 10^{-9}$--$1.6\times 10^{-8}$\,erg\,s$^{-1}$ per molecule,
at least two orders of magnitude larger than what was delivered by the
laser in the in vitro experiment~\cite{Nardecchia2018}.
Phonon condensation in vivo could therefore be activated on a far
shorter timescale than the minutes required in the laboratory.

\begin{figure}[h!] 
\centering
\includegraphics[scale=0.08,keepaspectratio=true]{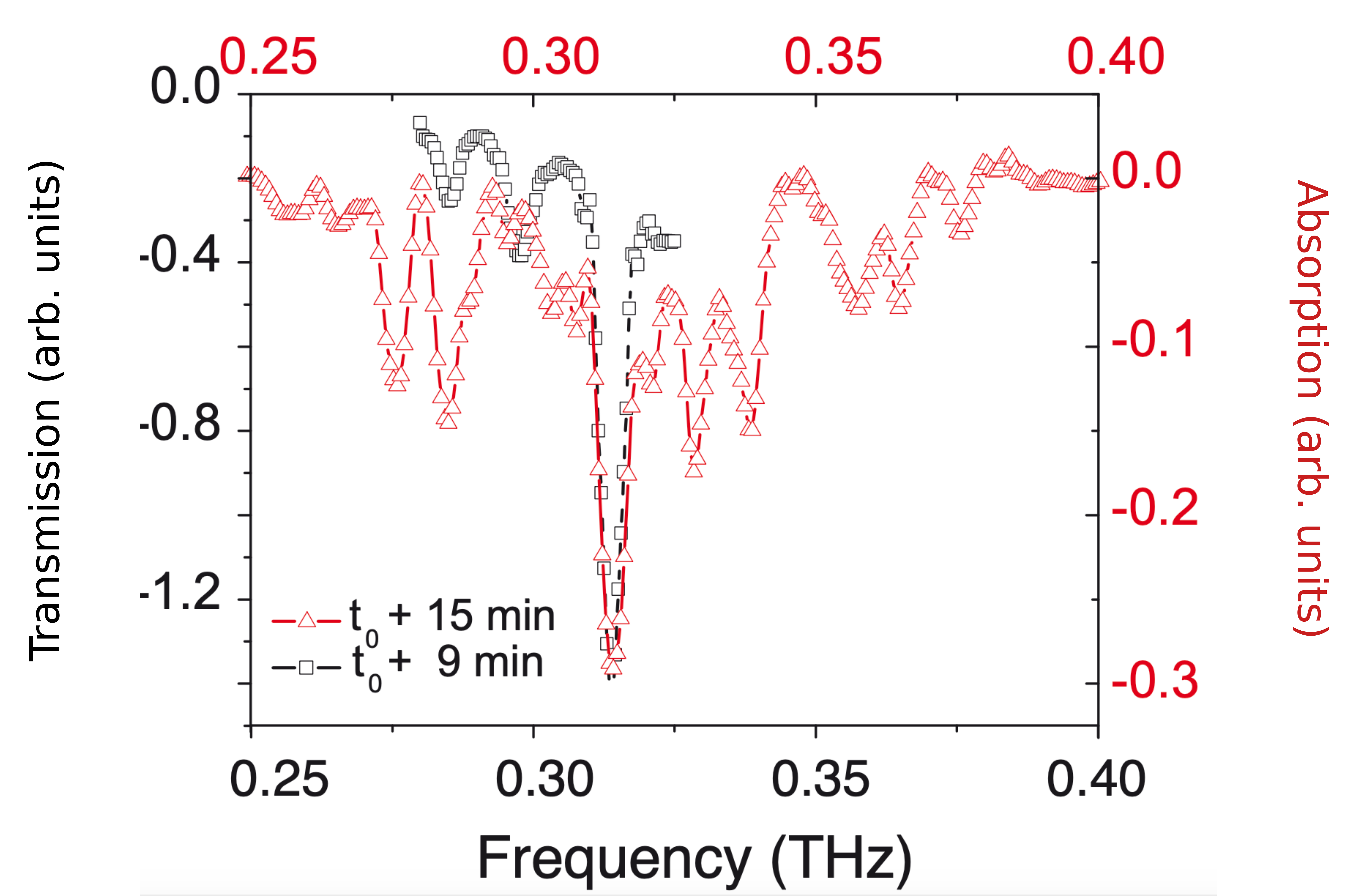} 
\caption{\sl Differential transmission and absorption spectra as functions of the frequency.  Comparison of the two normalized spectra for the longest illumination durations. From Ref.\protect\cite{Nardecchia2018}.} 
\label{fgr:2}
\end{figure}

\begin{figure}[h!] 
\centering
\includegraphics[scale=0.3,keepaspectratio=true]{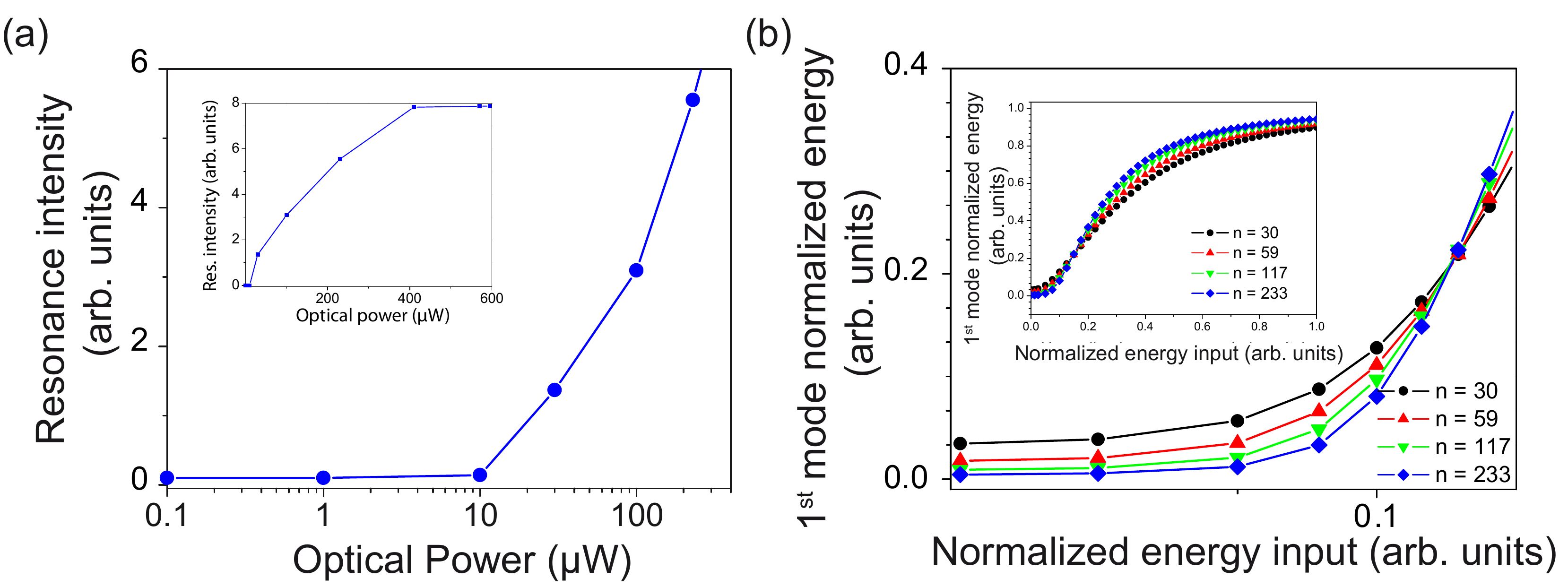} 
\caption{\sl Threshold-like behaviour of giant dipolar oscillations. (a) Intensity of the resonant peak measured at $0.314$ THz as a function of the optical laser power. (b) Normalized energy of the fundamental mode calculated as a function of the normalized source power. The different curves correspond to the reported numbers of normal modes of the BSA protein. Theory and experiment are in \textit{qualitative} agreement. From Ref.\protect\cite{Nardecchia2018}.}
\label{fgr:3}
\end{figure}

\subsection{Phonon condensation in R-PE}

R-phycoerythrin (R-PE), a hexameric light-harvesting protein of molecular
weight $240$\,kDa derived from red algae, provides a particularly clean
experimental system because it absorbs strongly at $488$\,nm through its
38 endogenous phycoerythrin and phycourobilin fluorochromes, without
requiring external labelling~\cite{Lechelon2022}.
Under laser illumination with a $488$\,nm source, a collective oscillation
mode appears at $\nu_0 = 71$\,GHz ($2.4$\,cm$^{-1}$), displaying both a
threshold in the energy input rate and a saturation of the oscillation
amplitude at high power, in full analogy with the BSA
observations~\cite{Nardecchia2018}.
The collective mode frequency is consistent with the lowest extension mode
of a torus of major radius $R = 37.5$\,\AA\ and minor radius $r = 30$\,\AA,
yielding a Young modulus $E \approx 5.3$\,GPa by inversion of the
Blevins formula~\cite{Lechelon2022}, a value that lies squarely between
those of myoglobin ($3.5$\,GPa) and BSA ($6.75$\,GPa), both predominantly
$\alpha$-helical proteins like R-PE.

Two collective extension modes appear at $71$\,GHz and $96$\,GHz, with
the frequency ratio $96/71 \approx 1.35$ matching the torus-mode
theoretical ratio $\sqrt{2} \approx 1.41$ within $4\%$.

The two-mode structure carries an important mechanistic signature of
phonon condensation.
The second mode at $96$\,GHz is observable only after the first mode at
$71$\,GHz has reached saturation~\cite{Lechelon2022}.
This is precisely what the nonlinear rate equations predict: the
condensation mechanism channels all injected energy into the fundamental
mode first; only once that mode is saturated does energy become available
to excite the next collective mode.
In the in vitro experiment, the onset time for the $71$\,GHz peak is
about $1$\,min (at 50\,mW), while approximately $15$\,min of pumping
is required to excite the $96$\,GHz peak.
This hierarchy of timescales is fully consistent with the condensation
picture and constitutes an additional internal consistency check on the
interpretation.
\begin{figure}[h!]
\centering
\includegraphics[scale=0.2,keepaspectratio=true,angle=0]{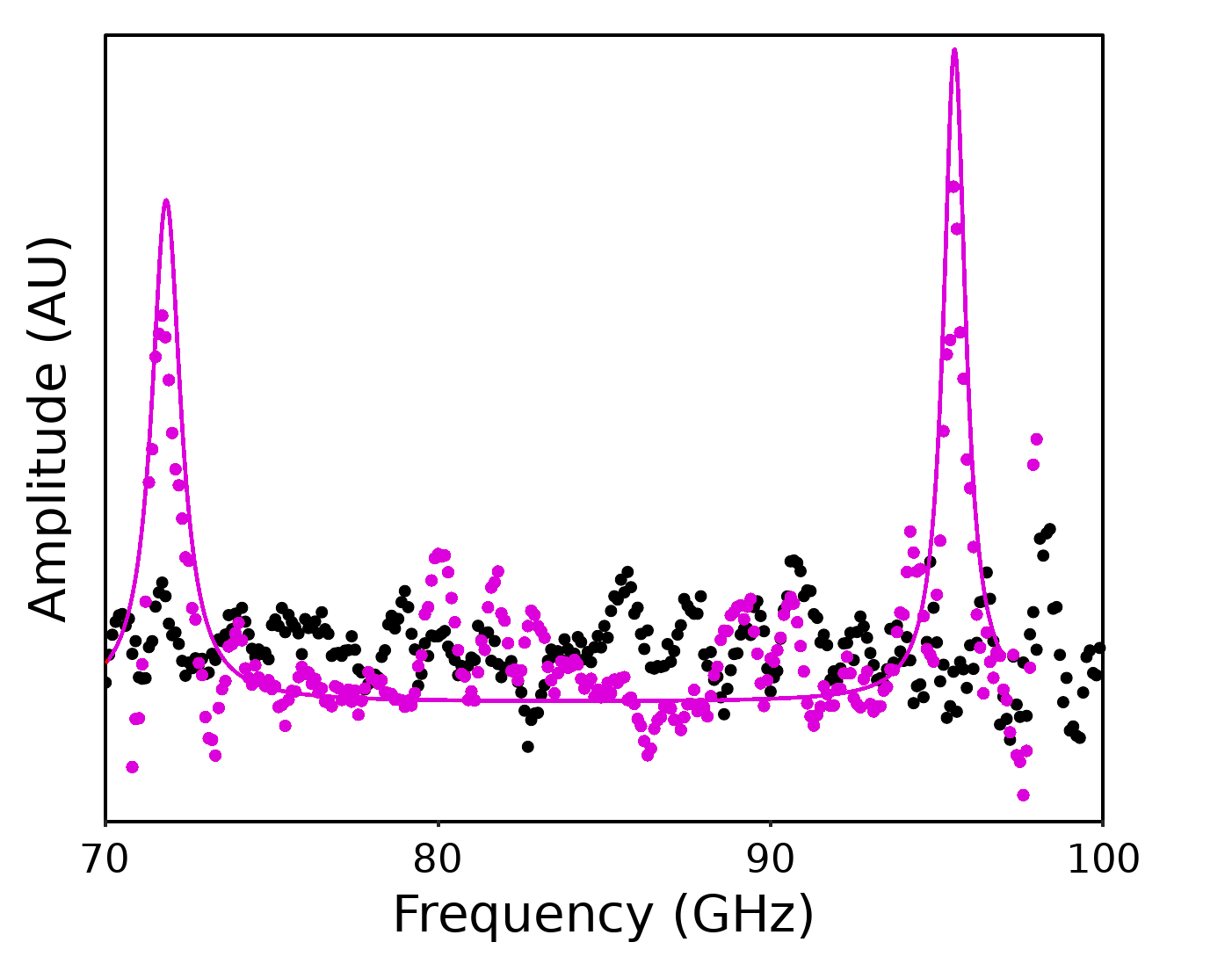}
\caption{{{\sl  R-PE coherent vibrational states.} Comparison of the R-PE in saline solution (purple) and saline solution without R-PE (black). Two collective extension modes of R-PE appear at 71 GHz and 96 GHz. Experimental data (full circles); Lorentz fit (solid line). From Ref.\protect\cite{Lechelon2022}.}} 
\label{RPEcoll-vibr}
\end{figure}
\subsection{Clustering phase transition}
The most striking macroscopic consequence of the activated electrodynamic
forces is a first-order clustering phase transition in R-PE solutions,
detected by fluorescence correlation spectroscopy (FCS)~\cite{Lechelon2022}.
The normalised diffusion coefficient $D/D_0$ measured by FCS is constant
across all concentrations at $50\,\mu$W (below threshold): purely Brownian
diffusion.
At $100\,\mu$W and $150\,\mu$W, $D/D_0$ drops by several orders of
magnitude at average intermolecular distances of $\approx 725$\,\AA\ and
$\approx 925$\,\AA\  respectively.
The transition point shifts to larger distances with increasing power
(stronger oscillations, stronger forces), and is fully reversible upon
switching off the laser.
\begin{figure}[h!]
%\centering
\includegraphics[scale=0.36,keepaspectratio=true]{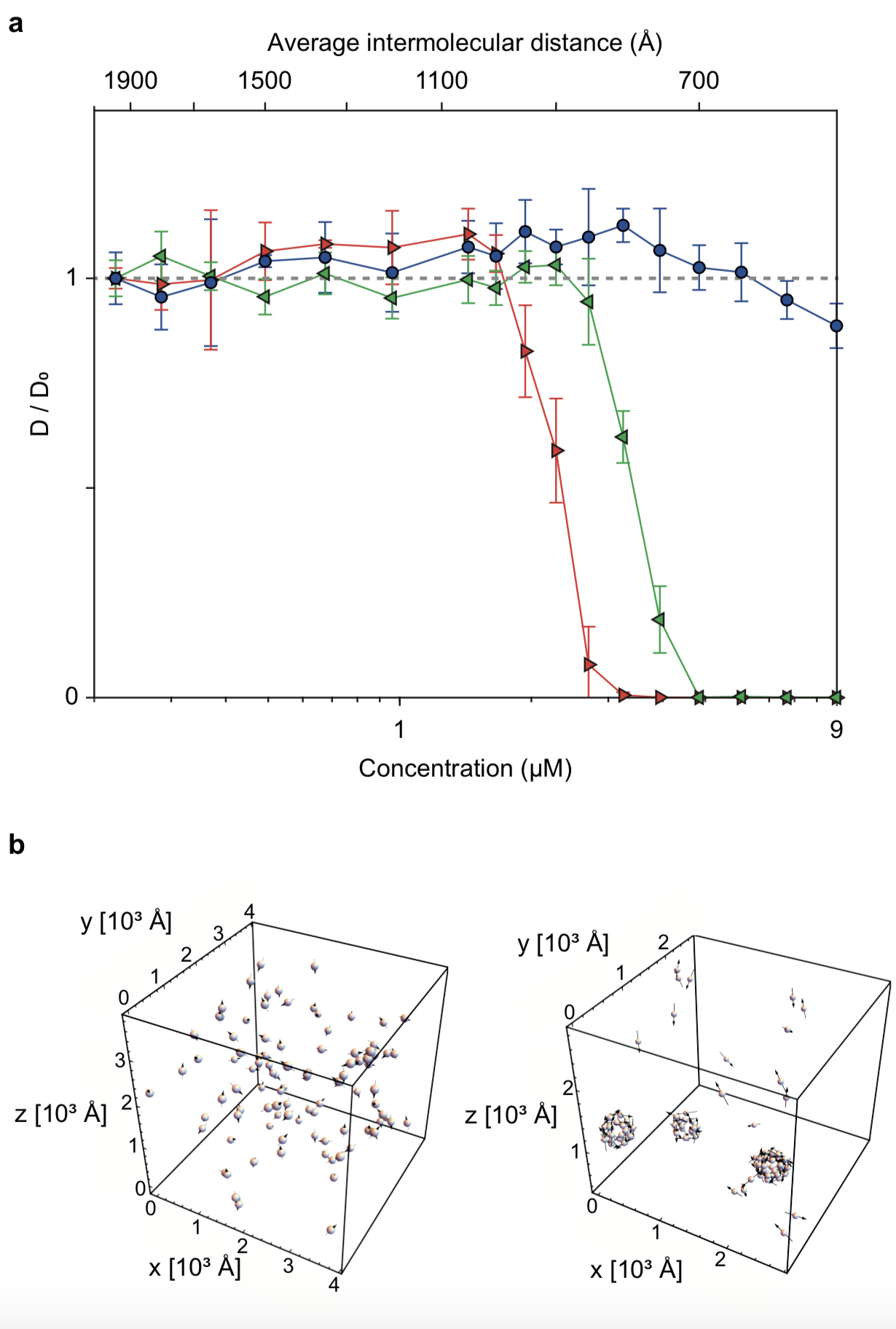} 
\includegraphics[scale=0.085,keepaspectratio=true]{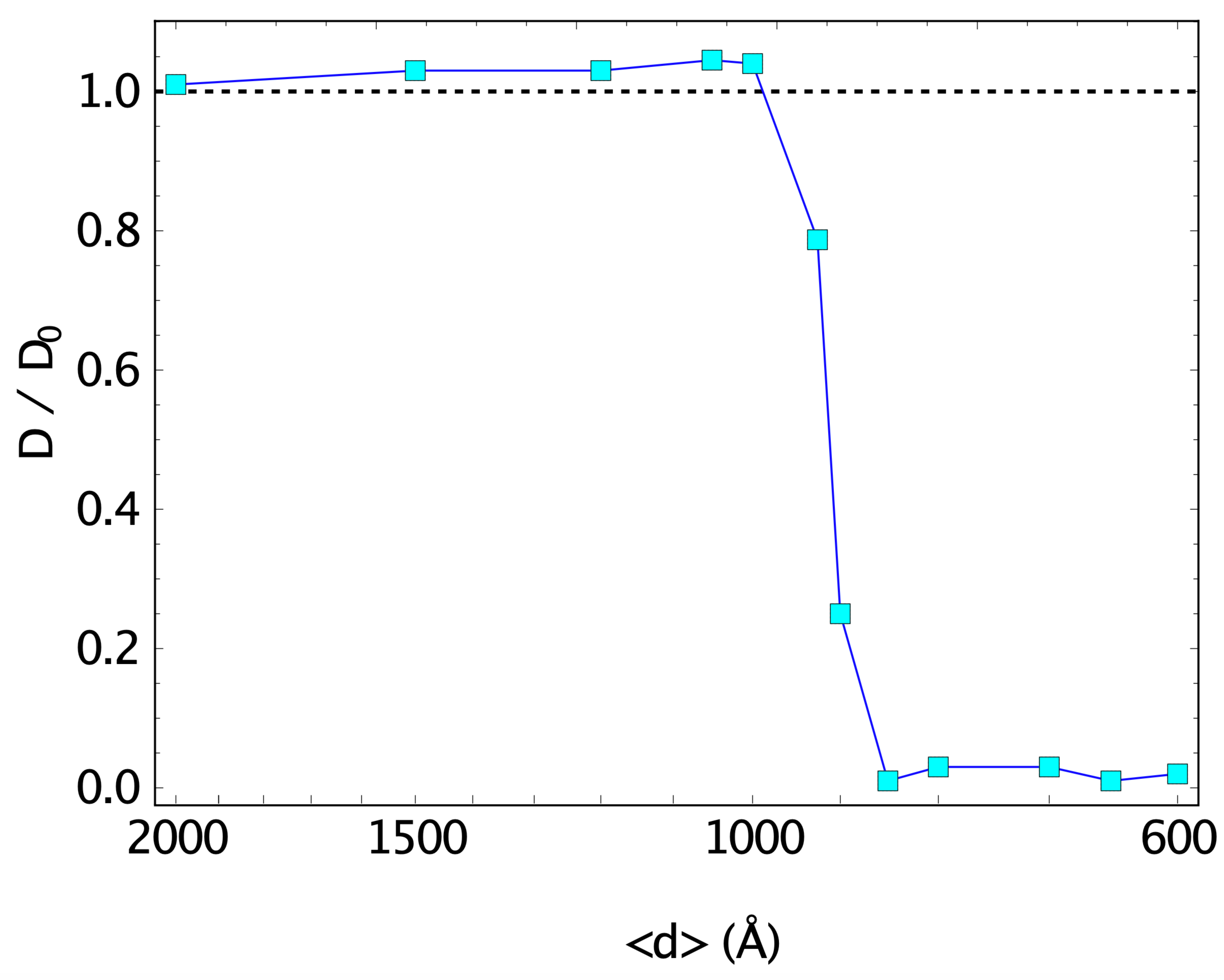}
\caption{{\sl Effect of protein concentration and laser power illumination on R-PE diffusion: Clustering phase transition.} (a) {Diffusion coefficients normalized to the Brownian $D_0$ values measured for each data series at 0.223 $\mu M$ ($\langle r\rangle \simeq 1950\ {\mathring{A}}$) and recorded at 50 $\mu W$ (blue circles), 100 $\mu W$ (green triangles), and 150 $\mu W$ (red triangles). Each point corresponds to the average of 5 independent experiments. (b) Clustering transition driven by long-range electrodynamic interparticle
forces, obtained from Monte Carlo simulations by varying the initial mean
separation $\langle r\rangle$. For $\langle r\rangle_{\rm init}\simeq 1000$\,\AA\ the system remains
in the dispersed phase ($D/D_0=1$, Brownian diffusion). For $\langle r\rangle_{\rm init}\simeq 950$\,\AA\ it switches to the
clustered phase ($D/D_0\ll 1$). Lower right panel: {Molecular Dynamics computation of self-diffusion coefficient $D/D_0$ (normalized to the Brownian value $D_0$) versus intermolecular average distance $\langle d \rangle$ for a system of particles in a cubic box interacting through long-range electrodynamic forces. From Ref.\protect\cite{Lechelon2022}.}}}
\label{RPE_clusterisation}
\end{figure}
The quantitative extent of the transition is remarkable: at an average
intermolecular distance of $600$\,\AA, the variance of fluorescence
fluctuations increases by more than five orders of magnitude compared
with the same quantity at $1950$\,\AA, and the measured diffusion time
$\tau_D$ jumps from $4.8\times 10^{-5}$\,s to $4.9$\,s, a factor of
$\sim 10^5$~\cite{Lechelon2022}.
These values cannot be explained by any trivial heating effect: a
thermal increase of diffusivity would shift $D_0$ uniformly and
monotonically with concentration, producing no step-like feature.
The molecular dynamics simulations with a $1/r^3$ interaction
potential reproduce the measured critical distance at $150\,\mu$W
quantitatively, providing independent theoretical support for the
experimental interpretation.
Monte Carlo computations confirm the formation of dense clusters below
the critical intermolecular distance~\cite{Lechelon2022}.
Direct visual evidence was obtained by confocal fluorescence
microscopy: clusters of R-PE molecules appear and grow when the laser
power exceeds the threshold, then rapidly dissolve when the laser is
switched back to a power below the threshold for activating collective
oscillations~\cite{Lechelon2022}.
This reversibility on a timescale of seconds rules out irreversible
photochemical aggregation and is fully consistent with the dynamical,
energy-dependent nature of the underlying ED forces.

The fact that BSA does not show the diffusion-based clustering signal
is itself informative and consistent with the theory.
As discussed above, BSA requires approximately
$10$\,min of optical pumping to activate a sharp collective oscillation.
In the FCS geometry, the transit time of each protein through the
confocal volume is far shorter than this activation time, so the
collective oscillations cannot be established during the measurement.
By contrast, R-PE, with its 38 endogenous fluorochromes and natural
light-harvesting efficiency, reaches the condensed state in less than
$3$\,s under comparable power density, making FCS measurement of its
clustering feasible~\cite{Lechelon2022}.
BSA is instead uniquely suited to the THz frequency-shift measurement,
which uses long-exposure illumination of a static drop of solution
rather than molecules in transit.
Thus the two proteins play complementary roles and together constitute
a self-consistent body of evidence.

These experimental outcomes are in quantitative agreement with the
theoretical prediction of a clustering transition driven by the resonant
$- 1/r^3$ electrodynamic interaction potential, activated out of thermal
equilibrium~\cite{Lechelon2022}.

\subsection{Frequency shifts: further evidence for long-range electrodynamic forces}

If resonant electrodynamic forces are active between collective dipoles
oscillating at frequency $\omega_0$, the vibration frequency of each
molecule shifts by $\Delta\nu \propto 1/\langle r\rangle^3 \propto \mathcal{C}$
(concentration), where $\langle r\rangle = \mathcal{C}^{-1/3}$ is the
average intermolecular distance~\cite{Preto2015,Lechelon2022}.
This prediction is confirmed for both R-PE (at $71$\,GHz) and BSA
(at $0.314$\,THz): the measured frequency shift varies linearly with
concentration, with a slope that increases with laser power (stronger
oscillation amplitude implies larger dipole moment and hence stronger
interaction).
Experiments were performed in $200$\,mM NaCl solution, ensuring
Debye screening of all electrostatic interactions.

For R-PE, frequency-shift measurements were performed at three laser
powers: $31.5$, $39.5$, and $50$\,mW.
In all cases the shift is linear in concentration, confirming the
$1/\langle r\rangle^3$ dependence over the range of average intermolecular
distances from roughly $700$ to $1100$\,\AA\ explored in the THz
experiments~\cite{Lechelon2022}.
The steepening of the slope with laser power is also quantitatively
consistent with theory: a larger oscillation amplitude generates a
larger oscillating dipole moment and hence a stronger electrodynamic
coupling.
The BSA frequency-shift measurement (at $40$\,mW) confirms the same
linear law at $0.314$\,THz, extending the evidence for long-range
electrodynamic forces to a structurally unrelated protein.
Crucially, the THz spectroscopy setup for these frequency-shift measurements
used a lower spatial density of laser light than the FCS diffusion
experiments, specifically to prevent a clustering transition during the
measurement: the ED forces were deliberately weakened to remain below
the threshold for cluster formation, so that the proteins kept diffusing
freely while still experiencing the frequency shift~\cite{Lechelon2022}.
This experimental design decouples the two observable signatures of
ED forces and rules out cross-contamination of the two measurements.

\begin{figure}[h!]
\centering
\includegraphics[scale=0.98,keepaspectratio=true]{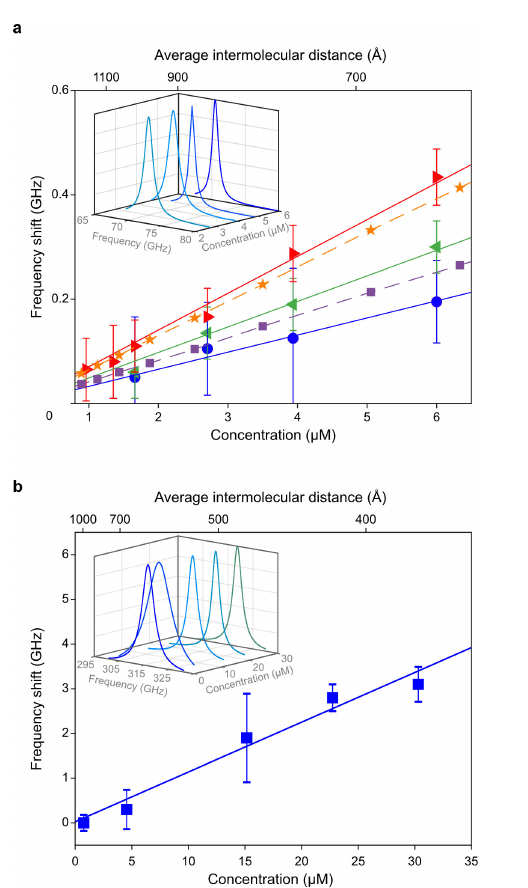} 
\caption{{\sl Frequency shifts of the intramolecular collective vibrations of R-PE and BSA at different concentrations.} {Measurements were performed at room temperature in aqueous solution with 200 mM of NaCl. Panel (a) refers to R-PE. The shift is relative to the reference frequency measured at the lowest  protein concentration. Measurements have been performed  at different powers of the laser: 31.5 mW (blue circles), 39.5 mW (green triangles), 50 mW (red triangles). Purple squares and orange stars refer to theoretical outcomes worked out with different values of molecular dipole moments (see Supplementary Materials). Panel (b) refers to BSA at a laser power of 40 mW.}~ Insets: The frequency shifts are measured through a Lorentz fitting of the experimental resonances. The different colors in the insets are just a visual help. From Ref.\protect\cite{Lechelon2022}. }
\label{f-shift}
\end{figure}

\newpage
%============================================================
\section{Discussion and Outlook}
\label{sec:outlook}
%============================================================
%\subsection{Unifying role of the TDVP}
A distinctive feature of the theoretical framework reviewed here is that
the TDVP serves as a single unifying instrument across different physical
scenarios.
In Sec.~\ref{sec:TDVP}, applied to product coherent states of bosonic
operators, it produces a classical Hamiltonian in action-angle variables
directly suitable for statistical (Liouville/KvN) analysis.
In Sec.~\ref{sec:DNA}, applied to the Davydov factorised ansatz of
electronic and phononic coherent states, it produces classical equations of
motion for electron probability amplitudes coupled to classical phonon degrees of
freedom, encoding the specific biomolecular sequence at the level of
site-dependent coupling constants.
In both cases the resulting dynamics is formally classical but directly
inherits the quantum structure of the original Hamiltonian, without invoking
any semiclassical approximation beyond the mean-field restriction to the
chosen ansatz manifold.

\subsection{Two complementary channels of selective long-range forces}

The review presents two physically distinct but mathematically parallel
channels through which selective long-range electrodynamic interactions can
be activated between biomolecules:

\emph{The vibrational (phonon) channel}: driven by metabolic or photonic
energy input, a macromolecule undergoes classical Fr\"ohlich condensation,
concentrating energy into a coherent collective oscillation.
The resulting giant oscillating dipole moment generates $1/r^3$
resonant forces, confirmed experimentally by THz spectroscopy and FCS.

\emph{The electronic channel}: electron--phonon excitation along a
specific DNA sequence or protein produces oscillating electron currents
whose Fourier spectrum carries the ``fingerprint'' of the biomolecular
sequence.
The cross-spectrum of two interacting molecules displays a sharp
co-resonance peak when and only when their sequences are cognate,
providing a sequence-specific electrodynamic interaction mechanism.

The two channels operate at different energy and frequency scales
(sub-THz vs.\ tens of THz) and are not mutually exclusive; in a living
cell they may well operate simultaneously and synergistically.

\subsection{Open questions}

\paragraph*{From light pumping to ATP.}
The experiments use laser excitation, which is a controlled but artificial
energy supply.
ATP hydrolysis provides an estimated power of $\sim 10^{-8}$--$10^{-7}$\,erg\,s$^{-1}$
per protein~\cite{Nardecchia2018} orders of magnitude larger than the
laser power available per protein in the experiments.
The position-space Hamiltonian framework~\cite{Preto2024}, via normal-mode
analysis of force fields, offers a path toward modelling ATP-coupled
condensation in atomistic detail.

\paragraph*{Role of water.}
The hydration shell plays an active role in both channels: it contributes to
the effective dipole moment in the phonon channel and provides a
Tavis--Cummings-like protective environment for electronic correlations in
the DNA--protein channel~\cite{Faraji2025,Kurian2018}.
A quantitative theory of the water contribution remains to be developed.

\paragraph*{In vivo evidence.}
Whether these phenomena operate in the complex, crowded, and dynamically
heterogeneous cellular environment is the central open question.
Selectivity of electrodynamic forces in a mixture of many molecular species,
the role of molecular crowding, and the competition between ED forces and
Brownian diffusion all require further investigation.

\paragraph*{Quantum corrections.}
The classical framework is quantitatively justified at sub-THz frequencies
and room temperature, but quantum fluctuations in the condensed
state~\cite{Zhang2019} and quantum coherence in the electronic channel may
be non-negligible and merit further study.

\paragraph*{Connection to biomolecular condensates.}
The first-order clustering transition driven by electrodynamic forces provides
a physically distinct mechanism for the formation of reversible biomolecular
assemblies~\cite{Banani2017}, complementing models based on intrinsically
disordered regions and multivalency.
Whether this mechanism is relevant to phase-separated cellular compartments
is an open and intriguing question.

\subsection{Concluding remarks}

The programme reviewed here demonstrates that classical Hamiltonian mechanics,
derived systematically from quantum models via the TDVP and implemented at
the level of both rate equations and direct Hamiltonian simulation, produces
rich and experimentally verifiable physics relevant to the organisation of
biochemical reactions in living matter.
Fr\"ohlich's half-century-old intuition has acquired rigorous Hamiltonian
underpinning, direct experimental support, and a new electronic dimension
through the DNA--protein co-resonance phenomenology.
The next challenge is to establish whether and how these phenomena operate
in the cellular environment, and what role they play in the extraordinary
efficiency of biological molecular recognition.

%============================================================
\begin{acknowledgments}
The authors wish to thank all the colleagues with whom the results reviewed here have been
obtained and have been reported in Refs.\cite{Nardecchia2018,Preto2015,Preto2024,Faraji2025,Faraji2021,Lechelon2022}, 
results that have been obtained partly within the project MOLINT funded by the Excellence Initiative 
of Aix-Marseille University - A*Midex, a French “Investissements d’Avenir” programme, and within the  project LINkS that has received funding from the European Union’s Horizon 2020 Research and Innovation Programme under grant agreement no. 964203 (FET-Open).
The authors used Claude (Anthropic) as an AI writing assistant only for light editing and polishing 
of the manuscript, not for conceptual development.
\end{acknowledgments}

%============================================================


\begin{thebibliography}{99}

\bibitem{Banani2017}
S.\ F.\ Banani, H.\ O.\ Lee, A.\ A.\ Hyman, and M.\ K.\ Rosen,
\textit{Biomolecular Condensates: Organizers of Cellular Biochemistry},
Nat.\ Rev.\ Mol.\ Cell Biol.\ \textbf{18}, 285 (2017).

\bibitem{Berry2018}
J.\ Berry, C.\ P.\ Brangwynne, and M.\ Haataja,
\textit{Physical Principles of Intracellular Organization via Active and
Passive Phase Transitions},
Rep.\ Prog.\ Phys.\ \textbf{81}, 046601 (2018).

\bibitem{Sweetlove2018}
L.\ J.\ Sweetlove and A.\ R.\ Fernie,
\textit{The Role of Dynamic Enzyme Assemblies and Substrate Channelling in
Metabolic Regulation},
Nat.\ Commun.\ \textbf{9}, 2136 (2018).

\bibitem{Frohlich1968}
H.\ Fr\"ohlich,
\textit{Long-Range Coherence and Energy Storage in Biological Systems},
Int.\ J.\ Quantum Chem.\ \textbf{2}, 641 (1968).

\bibitem{Frohlich1970}
H.\ Fr\"ohlich,
\textit{Long Range Coherence and the Action of Enzymes},
Nature (London) \textbf{228}, 1093 (1970).

\bibitem{Frohlich1972}
H.\ Fr\"ohlich,
\textit{Selective Long Range Dispersion Forces between Large Systems},
Phys.\ Lett.\ A \textbf{39}, 153 (1972).

\bibitem{Frohlich1977}
H.\ Fr\"ohlich,
\textit{Long-Range Coherence in Biological Systems},
Riv.\ Nuovo Cimento \textbf{7}, 399 (1977).

\bibitem{Wu1977}
T.\ M.\ Wu and S.\ Austin,
\textit{Bose Condensation in Biosystems},
Phys.\ Lett.\ A \textbf{64}, 151 (1977).

\bibitem{Wu1978}
T.\ M.\ Wu and S.\ Austin,
\textit{Bose-Einstein Condensation in Biological Systems},
J.\ Theor.\ Biol.\ \textbf{71}, 209 (1978).

\bibitem{Nardecchia2018}
I.\ Nardecchia, M.\ Pettini \textit{et al.},
\textit{Out-of-Equilibrium Collective Oscillation as Phonon Condensation
in a Model Protein},
Phys.\ Rev.\ X \textbf{8}, 031061 (2018).

\bibitem{Preto2015}
J.\ Preto, M.\ Pettini, and J.\ A.\ Tuszynski,
\textit{Possible Role of Electrodynamic Interactions in Long-Distance
Biomolecular Recognition},
Phys.\ Rev.\ E \textbf{91}, 052710 (2015).

\bibitem{Preto2024}
J. Preto, E. Floriani, V. Calandrini, G. Katona, M. Pettini, 
       \textit{ Hamiltonian model for energy condensation in classical systems: Relevance to proteins},
        Phys.Rev.Research \textbf{8}, L022016 (2026).

\bibitem{Faraji2025}
E.\ Faraji, V.\ Calandrini, P.\ Kurian, R.\ Franzosi, S.\ Mancini,
E.\ Floriani, G.\ Pettini, and M.\ Pettini,
\textit{Electrodynamic Forces Driving DNA--Protein Interactions at Large
Distances},
Front.\ Phys.\ \textbf{20}, 061200 (2025).

\bibitem{Bolterauer1999}
H.\ Bolterauer,
\textit{Elementary Arguments that the Wu--Austin Hamiltonian has no
Finite Ground State},
Bioelectrochem.\ Bioenerg.\ \textbf{48}, 301 (1999).

\bibitem{Kramer1980}
P.\ Kramer and M.\ Saraceno,
\textit{Geometry of the Time-Dependent Variational Principle in
Quantum Mechanics},
in \textit{Group Theoretical Methods in Physics} (Springer, 1980),
pp.\ 112--121.

\bibitem{Kramer2008}
P.\ Kramer,
\textit{A Review of the Time-Dependent Variational Principle},
J.\ Phys.\ Conf.\ Ser.\ \textbf{99}, 012009 (2008).

\bibitem{Jauslin2010}
H.-R.\ Jauslin and D.\ Sugny,
\textit{Dynamics of Mixed Classical-Quantum Systems, Geometric
Quantization and Coherent States},
in \textit{Mathematical Horizons for Quantum Physics},
IMS-NUS Lecture Notes, Vol.\ 20 (World Scientific, 2010).

\bibitem{Koopman1931}
B.\ O.\ Koopman,
\textit{Hamiltonian Systems and Transformation in Hilbert Space},
Proc.\ Natl.\ Acad.\ Sci.\ U.S.A.\ \textbf{17}, 315 (1931).

\bibitem{vonNeumann1932}
J.\ von Neumann,
\textit{Zur Operatorenmethode in der klassischen Mechanik},
Ann.\ Math.\ \textbf{33}, 587 (1932).

\bibitem{Reimers2009}
J.\ R.\ Reimers, L.\ K.\ McKemmish, R.\ H.\ McKenzie, A.\ E.\ Mark,
and N.\ S.\ Hush,
\textit{Weak, Strong, and Coherent Regimes of Fr\"ohlich Condensation
and Their Applications to Terahertz Medicine and Quantum Consciousness},
Proc.\ Natl.\ Acad.\ Sci.\ U.S.A.\ \textbf{106}, 4219 (2009).

\bibitem{Haken1975}
H.\ Haken,
\textit{Cooperative Phenomena in Systems Far from Thermal Equilibrium
and in Nonphysical Systems},
Rev.\ Mod.\ Phys.\ \textbf{47}, 67 (1975).

\bibitem{Cosic1994}
I.\ Cosic,
\textit{Macromolecular Bioactivity: Is it Resonant Interaction between
Macromolecules? Theory and Applications},
IEEE Trans.\ Biomed.\ Eng.\ \textbf{41}, 1101 (1994).

\bibitem{Veljkovic1985}
V.\ Veljkovic, I.\ Cosic, B.\ Dimitrijevic, and D.\ Lalovic,
\textit{Is it Possible to Analyze DNA and Protein Sequences by the Methods
of Digital Signal Processing?},
IEEE Trans.\ Biomed.\ Eng.\ \textbf{32}, 337 (1985).

\bibitem{Giese2000}
B.\ Giese,
\textit{Long-Distance Charge Transport in DNA: The Hopping Mechanism},
Acc.\ Chem.\ Res.\ \textbf{33}, 631 (2000).

\bibitem{GrayWinkler1996}
H.\ B.\ Gray and J.\ R.\ Winkler,
\textit{Electron Transfer in Proteins},
Annu.\ Rev.\ Biochem.\ \textbf{65}, 537 (1996).

\bibitem{Faraji2021}
E.\ Faraji, R.\ Franzosi, S.\ Mancini, and M.\ Pettini,
\textit{Transition between Random and Periodic Electron Currents on a DNA
Chain},
Int.\ J.\ Mol.\ Sci.\ \textbf{22}, 7361 (2021).

\bibitem{CosicBook}
I.\ Cosic,
\textit{The Resonant Recognition Model of Macromolecular Bioactivity:
Theory and Applications} (Birkh\"auser, Basel, 1997).

\bibitem{Kurian2018}
P.\ Kurian, A.\ Capolupo, T.\ J.\ A.\ Craddock, and G.\ Vitiello,
\textit{Water-Mediated Correlations in DNA-Enzyme Interactions},
Phys.\ Lett.\ A \textbf{382}, 33 (2018).

\bibitem{Ansari1985}
A.~Ansari, J.~Berendzen, S.~F.~Bowne, H.~Frauenfelder, I.~E.~Iben,
T.~B.~Sauke, E.~Shyamsunder, and R.~D.~Young,
Proc.\ Natl.\ Acad.\ Sci.\ U.S.A.\ \textbf{82}, 5000 (1985).

\bibitem{Lechelon2022}
M.\ Lechelon, M.\ Pettini \textit{et al.},
\textit{Experimental Evidence for Long-Distance Electrodynamic
Intermolecular Forces},
Sci.\ Adv.\ \textbf{8}, eabl5855 (2022).

\bibitem{Perticaroli2013}
S.\ Perticaroli, J.\ D.\ Nickels, G.\ Ehlers, H.\ O'Neill, Q.\ Zhang,
and A.\ P.\ Sokolov,
\textit{Secondary Structure and Rigidity in Model Proteins},
Soft Matter \textbf{9}, 9548 (2013).

\bibitem{Zhang2019}
Z.\ Zhang, G.\ S.\ Agarwal, and M.\ O.\ Scully,
\textit{Quantum Fluctuations in the Fr\"ohlich Condensate of Molecular
Vibrations Driven Far from Equilibrium},
Phys.\ Rev.\ Lett.\ \textbf{122}, 158101 (2019).

\end{thebibliography}
\end{document}